\documentclass{ws-ijmpd}
\usepackage{bm}
\usepackage{braket}
\usepackage[super,compress]{cite}
\usepackage{epsfig}
\usepackage{colordvi}
\usepackage{psfrag}
\usepackage{color}

\usepackage{mathrsfs}

\newcommand{\Lagr}{\mathcal{L}}
\newcommand{\G}{\mathcal{G}}

\usepackage{hyperref}
\usepackage{epstopdf}
\usepackage{times}

\hypersetup{
                    colorlinks=true,        
                    linkcolor=blue,         
                    citecolor=magenta,      
}

\begin{document}

\markboth{Salvatore Capozziello}{Francesco Bajardi}

%
\catchline{}{}{}{}{}
%

\title{Non-Local Gravity Cosmology: an Overview}


\author{Salvatore Capozziello}
\address{Department of Physics ``E. Pancini'', University of Naples ``Federico II'', Naples, Italy. \\ INFN Sez. di Napoli, Compl. Univ. di Monte S. Angelo, Edificio G, Via	Cinthia, I-80126, Naples, Italy. \\ Scuola Superiore Meridionale, Largo San Marcellino 10, I-80138, Naples, Italy. \\ Laboratory of Theoretical Cosmology, Tomsk State University of Control Systems and Radioelectronics (TUSUR), 634050 Tomsk, Russia. \\
Department of Mathematics, Faculty of Civil Engineering,
VSB-Technical University of Ostrava, Ludvika Podeste 1875/17, 
708 00 Ostrava-Poruba, Czech Republic.\\
\email{capozziello@na.infn.it}}

\author{Francesco Bajardi}
\address{Department of Physics ``E. Pancini'', University of Naples ``Federico II'', Naples, Italy. \\ INFN Sez. di Napoli, Compl. Univ. di Monte S. Angelo, Edificio G, Via Cinthia, I-80126, Naples, Italy. \\
\email{francesco.bajardi@unina.it}}

\maketitle

\begin{history}
\received{Day Month Year}
\revised{Day Month Year}
\end{history}


\begin{abstract}
We discuss some  main aspects of  theories of gravity containing non-local terms in view of cosmological applications. In particular, we consider various extensions of General Relativity   based on  geometrical invariants as  $f(R, \Box^{-1} R)$, $f(\G, \Box^{-1} \G)$ and $f(T, \Box^{-1} T)$ gravity  where $R$ is the Ricci curvature scalar,  $\G$ is the Gauss-Bonnet topological invariant,  $T$ the torsion scalar and the operator $\Box^{-1}$ gives rise to non-locality. After selecting their functional form by using Noether Symmetries, we find out exact solutions in a cosmological background. It is possible to reduce the dynamics of selected models and  to find analytic solutions for the equations of motion. As a general feature of the approach, it is possible to address the accelerated expansion of the Hubble flow at various epochs, in particular the dark energy issues, by taking into account non-locality corrections to the gravitational Lagrangian.  On the other hand, it is possible to search for gravitational non-local effects also at astrophysical scales. In this perspective,  we search for symmetries of $f(R, \Box^{-1} R)$ gravity also  in a spherically symmetric background and constrain the free parameters, Specifically, by taking into account the S2 star orbiting around the Galactic  Centre SgrA$^*$, it is possible to study how non-locality affects stellar orbits around such a massive self-gravitating object.
\end{abstract}

\ccode{PACS numbers: 98.80.-k, 95.35.+d, 95.36.+x}

\maketitle


\section{Introduction}

Quantum Mechanics is the most revolutionary theory in Physics formulated in  the last century. It  opened the doors to a completely new vision of Nature at any scale. The determinism of Classical Mechanics was replaced by a probabilistic view of small-scale phenomena, which seemed to be the only way to fit all the experimental results. As we gained a theory capable of describing almost all the evidences provided by the Quantum World, we lost the capability to exactly predict the time evolution of quantum systems. Soon after, Quantum Field Theory (QFT) arose with the purpose to describe all the fundamental interactions under the same standard. It was soon clear that this prescription could not be applied to the gravitational interaction. Indeed, as Quantum Mechanics is probabilistic by nature, gravity is, in turn, described by the Einstein General Relativity (GR), where non-local interactions are not allowed. 
So far, a theory capable of describing both the large-scale structure and the Ultraviolet (UV) scale  is still missing. Moreover, neither QFT nor GR hold at the Planck scale, where a new physics is probably needed. 
On the one hand, despite all the experimental confirmations of Quantum Mechanics, we still miss its deep meaning; on the other hand, although GR is mathematically consistent and well developed, it provides some incompatibilities both at   large and small-scale regimes. Any attempt to merge the formalism of GR with that of QFT  failed till now. Even though QFT in curved spacetime addresses several evidences provided by the small-scales observations (such as the Hawking Radiation, the Unruh effect or the cosmic inflation) it suffers several shortcomings. Indeed, it turns out that GR can be renormalized up to the second loop level \cite{Goroff:1985th}, which means that incurable divergences arise once adapting the same scheme as QFT to gravity. In addiction, unlike the other fundamental interactions, GR cannot be treated under the standard of Yang-Mills theories, due to the lack of a corresponding Hilbert space and a probabilistic interpretation of the gravitational wave function. For these reasons, a coherent and self-consistent theory of Quantum Gravity is one of the most studied topic nowadays \cite{Horava:2009uw, Stelle:1976gc, DeWitt:1967ub, Engle:2007uq, Alexandrov:2002br, Maldacena:1997re, Nilles:1983ge}. In the last few years, the quantum formalism was adapted to cosmology, where the dynamics can be reduced considering a minisuperspace of  variables. It represents a "toy model" which yields several important results towards understanding the early-stages of our Universe \cite{Hartle:1983ai, Hawking:1983hj, Halliwell:1984eu, Vilenkin:1984wp, Bajardi:2020fxh}. However, the so-called Quantum Cosmology is far from being a comprehensive and self-consistent Quantum Gavity theory.

At the astrophysical scales, GR soon obtained a great success after the observations of the light deflection, followed by the Radar Echo Delay and the exact estimation of the perihelion of Mercury precession in its orbit around the Sun. The recent discoveries of gravitational waves and black holes confirm the relativistic picture of astrophysical phenomena.

The application of GR to a homogeneous and isotropic spacetime led to trace a self-consistent cosmic history ranging from the Big Bang to the Dust Matter Dominated Era. Despite all these successful results,  some shortcomings emerged in the recent years. For instance,  GR is not able to predict the mass-radius profile of compact objects, or the speed of the farthest stars orbiting around the center of a given galaxy, which is experimentally lower than that theoretically expected \cite{Babocock, Bosma:1981zz}). To  fix the latter issue, the missing matter was addressed considering  a hypothetical fluid with zero pressure, called \emph{Dark Matter}, which should account for the 26$\%$ of the Universe, but which has never  experimentally detected, at fundamental level, in the form of some new particles. On the other hand, the cosmological constant $\Lambda$ was introduced to explain the today observed accelerated expansion of the Universe, dubbed \emph{Dark Energy}. The latter is supposed to represent more or less  68$\%$ of the whole energy-matter content of the universe. The accelerated expansion cannot be predicted by GR without invoking the presence of Dark Energy, as the Galaxy Rotation Curve cannot be fitted without Dark Matter. However, even introducing the cosmological constant in the field equations, at  quantum level, there is a discrepancy of 120 orders of magnitude between the theoretical value of $\Lambda$ and the today observed  experimental  one.

For all these reasons, in the last years, several new theories have been developed, with the aim to address some of these issues. In most cases the starting point is a modification of the Einstein-Hilbert gravitational action, including other curvature invariants or the coupling between geometry and scalar fields \cite{Clifton:2011jh, Nojiri:2017ncd, Joyce:2014kja, Capozziello:2019klx, Capozziello:2011et, Capozziello:2019cav, Bajardi:2020xfj}. The simplest extension is the $f(R)$ gravity \cite{Cap, Nojiri:2006gh, DeFelice:2010aj, Sotiriou:2008rp, Starobinsky:2007hu}, whose action contains a function of the scalar curvature, and whose field equations are of  fourth order with respect to the metric. For some form of the function, the theory can fit the Galaxy Rotation Curve without Dark Matter, or the exponential expansion of the universe without Dark Energy \cite{Cap,Copeland:2006wr, Nojiri:2006ri, Bamba:2012cp}. Among $f(R)$ gravity models, one of the most famous is the Starobinsky model, whose action differs from the Einstein-Hilbert one for a quadratic term in the scalar curvature \cite{Starobinsky:1980te}. The theory well describes the cosmic inflation according to the  experiment data. Other possible modifications of GR deal with  non-minimal coupling with one or more dynamical scalar fields. Such a class of  theories was firstly proposed  by Linde and Guth \cite{Guth:1980zm, Guth:1982ec, Linde:1981mu}. Moreover, it turns out that scalar-tensor theories can be recast into higher-order theories by means of conformal transformations. A part from $f(R)$ gravity, fourth-order theories can be constructed by means of several curvature invariants, such as $R_{\mu \nu} R^{\mu \nu}$ or $R_{\mu \nu \rho \sigma} R^{\mu \nu \rho \sigma}$. In particular, the  combination
\begin{equation}
\mathcal{G} \equiv R^2 - 4 R_{\mu \nu} R^{\mu \nu} + R_{\mu \nu p \sigma} R^{\mu \nu p \sigma},
\end{equation}
called \emph{Gauss-Bonnet} scalar, plays an important role in modified theories of gravity. It represents a topological surface term \cite{Glavan:2019inb, Duan:2006zt}, whose integral over the four-dimensional manifold provides the Euler Characteristic \cite{Alty:1994xj}. Although it does not give any contribution to the field equations, a function of $\G$ turns out to be non-trivial in more than three dimensions; therefore, $f(\G)$ theory of gravity can be considered in cosmological or spherically symmetric backgrounds to address GR inconsistencies \cite{Bajardi:2020osh, Bajardi:2019zzs}. The introduction of higher-order curvature invariants into the action is also motivated by the semiclassical approach of QFT in curved spacetime, where gravity is treated from a classical point of view, and the geometry is coupled to a quantum energy-momentum tensor of the ordinary matter. The one-loop effective action of GR exhibits high-order invariants, such that the effective matter Lagrangian contains new UV divergent terms proportional to $R^2$ and $R^{\mu \nu} R_{\mu \nu}$ \cite{Birrell:book, Adams:1990pn, Cotsakis:2006zn, Amendola:1993bg}.

Another strict hypothesis of GR is the symmetry of Christoffel connection to describe the geodesic structure. In the case of Levi-Civita, where affine connections are derived by the combination of metric and its derivatives,   the dynamics  is described by curvature and the coincidence of metric structure with geodesic structure is guaranteed by the Equivalence Principle. Including the anti-symmetric part of the connection, it is possible to define a non-vanishing rank-3 tensor of the form
\begin{equation}
T^\alpha_{\mu \nu} \equiv \Gamma^\alpha_{\mu \nu} -  \Gamma^\alpha_{\nu \mu}.
\end{equation}
In this formalism, torsion arises as a fundamental field, at the same level as curvature. The rank-3 tensor $T^\alpha_{\mu \nu}$ is the \emph{Torsion Tensor}, whose main aspects are pointed out in Sec. \ref{ETGs}. Starting from the torsion tensor, a specific contraction of the indexes permits to define the \emph{Torsion Scalar} $T$. The theory which describes the spacetime dynamics  only with torsion and without curvature is called \emph{Teleparallel Equivalent to General Relativity} (TEGR). The teleparallel Lagrangian differs from Einstein--Hilbert one only for a four-divergence, which means that TEGR and GR are dynamically equivalent \cite{Hammond:2002rm, Maluf:2013gaa, Aldrovandi:2015wfa, Aldrovandi:2013wha}.
However, the other fundamental differences with respect to GR are that tetrads $h^a_\mu$ constitute  the fundamental objects, instead of metric,  describing the gravitational field  and,  starting from them, it is possible to define the Weitzenb\"ock connection, representing the affine connection in TEGR instead of the Levi-Civita one, as we will see below. 
As a natural extension of TEGR,  functions of the torsion scalar can be considered into the action, with the aim to solve the  large-scale structure and cosmic acceleration problems arising in the context of GR \cite{Cai:2015emx, Li:2010cg, Wu:2010mn, Capozziello:2011hj, Boehmer:2011gw, Wu:2010xk, Iorio:2012cm, Bahamonde:2019zea, Bahamonde:2016grb}. The great advantage of TEGR is due to the description of the gravitational interaction in the locally flat spacetime; in this way, TEGR can be recast as a Gauge Theory of the translation group \cite{Aldrovandi:2015wfa, Aldrovandi:2013wha}. 

Because of the general form of the above mentioned alternatives to GR, a selection criterion aimed at constraining the starting action is needed. Testing a theory of gravity through observational  cosmology is a good starting point in order to select some consistent theories and discard some others. However,  extending GR leads to hardly solvable field equations, so that the given theory ends up losing any predictive power. In order to reduce the dynamics and find suitable equations, it is possible to  adopt a selection criterion based on Noether symmetries, discussed in detail in \ref{noeth}. In few words, the existence of Noether symmetries for a given class of models allows to select viable Lagrangians whose dynamics is reducible and presents conserved quantities. These one are first integrals  of motion which allow to solve the related dynamical system. It is interesting to point out that the existence of Noether symmetries is always related to physically meaningful models \cite{Rugg}.

This review paper is organized as follows: in Sec. \ref{non-local-sect} we overview some  aspects of non-locality in Physics. In Sec. \ref{ETGs} we introduce local and non-local theories of gravity, briefly summarizing their main properties. Secs. \ref{BoxR}, \ref{BoxG} and \ref{BoxT} are devoted to the  study the Lagrangian formalism of some modified non-local actions and the Noether Symmetry Approach is thus applied in order to solve dynamics. Specifically, in Sec. \ref{BoxR}, curvature based   non-local actions are considered, while in Secs. \ref{BoxG} and \ref{BoxT} the Gauss-Bonnet scalar and the torsion scalar are  included. In Sec. \ref{sectS2}, a non-local function of the Ricci scalar, selected by Noether's theorem, is adopted to construct a spherically symmetric metric. The aim is reproducing the motion of stars around the Galactic Centre. We constrain the parameters of the theory adopting the observational data  coming from the  S2 star orbit. Finally, in Sec. \ref{Concl}, we summarize the main results and conclude  with  some considerations on new perspectives.

\section{Non-Locality in Physics} \label{non-local-sect}
 Non-locality naturally emerges in Quantum Physics and can be considered  one of the main issues arising in the attempt to merge the formalism of QFT with that of GR. As a matter of facts, while classical theories are \emph{local theories}, Quantum Mechanics is kinematically non-local. It is important to distinguish between kinematical and dynamical locality/non-locality: while the former refers to the states describing a theory, the latter relies to the interactions, so that the dynamical locality (non-locality) is only due to the local (non-local) form of the action. 

In general, any transcendental function of fields can be represented by the integral kernels of differential operators, namely the operator $\Box^{-1} \equiv (g_{\mu \nu} D^\mu D^\nu)^{-1}$ (with $D^\mu$ being the covariant derivative), which accounts for long-range non-local effects. It can be written in terms of the associated Green function $G(x,x')$ as:
\begin{equation*}
\square^{-1}\phi\,(x)\equiv\int d^4x' \, G(x,x')\phi(x')\,.
\end{equation*}

Any classical theory can be localized at any time of its evolution, so that it is possible to perform a measure with arbitrary precision without perturbing the state. Thus, any classical field theory is kinematically local.

On the contrary, Quantum Mechanics is non-local by construction.  According to the Heisenberg Principle, a given particle cannot be localized with arbitrary precision. Due to this property, a particle starting its motion from a given position $x_1$ and moving towards a final position $x_2$ follows any possible path linking the two points. Unlike what happens in Classical Mechanics, where the actual path between two fixed endpoints is uniquely selected by the initial conditions, in Quantum Mechanics all paths are simultaneously allowed.  

It is worth pointing out that the kinematical non-locality of the theory does not necessarily imply dynamical non-locality: in order for a theory to be dynamically non-local, the action must have a non-local form as well.  Another manifestation of non-locality in Quantum Mechanics is given by the entanglement, which exhibits a kinematical non-locality due to the interaction at a distance between particles.

In general, in the formalism of QFT, all the fundamental interactions exhibit dynamical non-locality as soon as their one-loop effective actions are considered \cite{Barvinsky:2014lja}. This is the case, \emph{e.g}, of the Euler-Heisenberg Lagrangian:
\begin{equation}
\mathcal{L}_{EH}=-\frac{1}{4}\mathcal{F}^2-\frac{e^2}{32\pi^2}\int_0^{\infty}\frac{ds}{s}\,e^{i\varepsilon s}%
                            e^{-m^2s}\biggl[\frac{\textup{Re} \cosh(esX)}{\textup{Im} \cosh(esX)}\mathcal{F}_{\mu\nu}\widetilde{\mathcal{F}}^{\mu\nu}-%
                            \frac{4}{e^2s^2}-\frac{2}{3}\mathcal{F}^2\biggr] \,,
                            \label{LagrE-Heisen}
\end{equation}
where $\mathcal{F}_{\mu \nu}$ is the electromagnetic tensor, defined through the potential $A_\mu$ as $\mathcal{F}_{\mu \nu} = \partial_\mu A_\nu - \partial_\nu A_\mu$. The scalars $\mathcal{F}$ and $X$ are defined respectively as $\mathcal{F} =  \frac{1}{2} \left(|\textbf{E}|^2 - |\textbf{B}|^2 \right)$ and $X =\mathcal{F}+ i \, \textbf{E} \cdot \textbf{B}$, while the constant $e$ and $m$ are the charge and the mass, respectively. The above Lagrangian is the renormalized one-loop effective Lagrangian which arises after integrating out a massive fermion from the quantum electrodynamics full Lagrangian.  Eq. \eqref{LagrE-Heisen} is of particular interest, since the non-locality is due to the intrinsic non-local nature of an integration, which can be understood as the inverse of a differential operator.

Another example is given by the low-energy limit of the Yukawa theory with a massive scalar field $\phi$, whose effective Lagrangian reads
\begin{equation*}
\mathcal{L}_Y=i \bar{\psi}/\!\!\!\partial\psi - \frac{1}{2} \phi(\square+m^2)\phi+\lambda\,\phi\bar{\psi}\psi %
\quad \to \quad \mathcal{L}_{eff}=i \bar{\psi}/\!\!\!\partial\psi+\frac{\lambda^2}{2}\bar{\psi}\psi(\square+m^2)^{-1}\bar{\psi}\psi \,.
\end{equation*}
The non-locality here is provided by the operator $(\square+m^2)^{-1}$. After performing a straightforward Taylor expansion, the Lagrangian can be written as
\begin{equation*}
\mathcal{L}_{eff}=i \bar{\psi}/\!\!\!\partial\psi+\frac{\lambda^2}{2m^2}\bar{\psi}\psi\bar{\psi}\psi-%
                            \frac{\lambda^2}{2m^4}\bar{\psi}\psi\square\bar{\psi}\psi+\ldots \,\, ,
\end{equation*}
which coincides with the 4-Fermi theory when the scalar field $\phi$ accounts for the W and Z bosons.  In the next section we focus on the application of non-locality to gravity theories classifying  them in two main categories.  
\section{Modified and Extended Theories of Gravity} \label{ETGs}
\subsection{Local Theories of Gravity}

As briefly discussed in the introduction, several assumptions of GR are not motivated by experimental observations. For instance, the action is supposed to be linearly dependent on the scalar curvature, so that second-order field equations occur from its variation with respect to the metric, in close analogy with Maxwell equations. Moreover, the affine connection is assumed to be symmetric with respect to the lowest indexes, yielding a torsionless spacetime. The reason of this assumption is the Equivalence Principle which is fully compatible with the Levi-Civita connection. On the other hand, in the Einstein-Cartan formalism,  the antisymmetric part of the connection is included in the curvature tensor and the spacetime turns out to be described by  both curvature and torsion \cite{Hehl:1976kj}. 


If one relaxes the constraint of second-order field equations, the Einstein-Hilbert Lagrangian can be extended by introducing other curvature invariant terms. One of the most studied extended Lagrangian is the $f(R)$ gravity, given by the Lagrangian
\begin{equation}
\Lagr = \frac{\sqrt{-g}}{2 \kappa} \,\left[ f(R) + \Lagr^{(m)}\right],
\end{equation}
where $f(R)$ is a generic function of the Ricci scalar, $\Lagr^{(m)}$ is the matter Lagrangian and $\kappa$ is the gravitational coupling defined as  $\kappa = 8 \pi G_N$, with $G_N$ being the Newton constant. The above Lagrangian leads to the following fourth-order field equations with respect to the metric tensor:
\begin{equation}
\displaystyle f_R(R) R_{\mu \nu }-{\frac {1}{2}}f(R)g_{\mu \nu }+\left[g_{\mu \nu }\Box -D_{\mu }D_{\nu }\right] f_R(R)= \kappa T_{\mu \nu },
\end{equation}
where $f_R$ is the first derivative of $f(R)$ with respect to $R$ and $T_{\mu \nu}$ is the energy-momentum tensor of matter fields. These theories, under given limits, are able to solve most of the inconsistencies provided by GR at the astrophysical scales, but Einstein's gravity is restored as a particular case of them \cite{Capozziello:2006dj, Capozziello:2007eu, Capozziello:2007ms, Capozziello:2007id, Capozziello:2008fn}. 

The higher-order derivatives of the metric can be recast as an effective geometric stress-energy tensor, which can mimic the role of Dark Matter and Dark Energy \cite{Mantica}. A famous example, mostly used to describe the early-universe evolution, is given by Starobinsky's gravity \cite{Starobinsky:1980te}, whose action reads
\begin{equation*}
\label{ETG:11}
\mathcal{S}_{Starobinsky}=\frac{1}{2 \kappa}\int \!d^4x \sqrt{-g}\,\bigl[R+\alpha R^2\bigr]+\mathcal{S}^{(m)} ,
\end{equation*}
where $\mathcal{S}^{(m)}$ is the ordinary matter action. The quadratic term $R^2$ can give rise to the  acceleration   of the early universe giving rise to an inflationary behavior. 

Other modified actions can be constructed by including  combinations of the Riemann,  Ricci and Weyl tensors. In this framework, the most general action containing the scalars $R$, $R^{\mu \nu}R_{\mu \nu}$ and $R^{\mu \nu p \sigma} R_{\mu \nu p \sigma}$ is
\begin{equation}
S = \int \sqrt{-g} f(R, R^{\mu \nu}R_{\mu \nu}, R^{\mu \nu p \sigma} R_{\mu \nu p \sigma}) \, d^4x.
\label{f(R,P,Q)}
\end{equation}
By defining
\begin{equation}
P \equiv g^{\mu p} R^{\nu \sigma} R_{\mu \nu p \sigma}, \,\,\,\,\,\,\,\,\,\,\,\,\,\,\, Q \equiv R^{\mu \nu p \sigma} R_{\mu \nu p \sigma},
\end{equation}
the variation of the action \eqref{f(R,P,Q)} provides:
\begin{eqnarray}
&& f_R(R,P,Q) G_{\mu \nu} = \left[\frac{1}{2} g_{\mu \nu} f(R,P,Q) - R f_R(R,P,Q) \right]   \nonumber
\\
&-& \left(g_{\mu \nu} \Box - D_\mu D_\nu \right)f_R(R,P,Q)  \nonumber
\\
&-& 2 \left[ f_P(R,P,Q) R^\alpha_\mu R_{\alpha \nu} + f_Q(R,P,Q) R_{p \sigma \alpha \mu} R^{p \sigma \alpha}_{\;\;\;\;\;\; \nu} \right] \nonumber
\\
&-& g_{\mu \nu} D_p D_\sigma \left[f_P(R,P,Q) R^{p \sigma}\right] - \Box \left[f_P(R,P,Q) R_{\mu \nu} \right]   \nonumber
\\
&+& 2 D_\sigma D_p \left[f_P(R,P,Q) R^p_{\left\{ \mu \right.} \delta^\sigma_{ \nu \left. \right\}} + 2 f_Q(R,P,Q) R^{p\;\;\;\;\; \sigma}_{\; \left\{ \mu \nu\right\}} \right] 
\label{field equations f(R,P,Q)}\,,
\end{eqnarray}
where $ \{ \} $ is the anti-commutator. The action \eqref{f(R,P,Q)} admits as a subcase the so-called Stelle's gravity \cite{Stelle:1976gc, Stelle:1977ry}, namely
\begin{equation*}
S_{Stelle}=\frac{1}{2 \kappa} \int \!d^4x \sqrt{-g}\, \bigl[R+\alpha R^2+\beta R^{\mu\nu}R_{\mu\nu}\bigr] +\mathcal{S}^{(m)} \,,
\end{equation*}
which is renormalizable at the quantum level. By properly combining $P,Q$ and $R^2$, it is possible to construct a topological surface term, called the Gauss-Bonnet invariant:
\begin{equation}
\G = R^2 -4R^{\mu \nu}R_{\mu \nu} + R^{\mu \nu p \sigma} R_{\mu \nu p \sigma}.
\end{equation}
Being a surface term, the integration over the manifold results into a topological invariant; specifically, according to the generalized Gauss-Bonnet theorem, in four dimensions $\G$ is the Euler density, while its integration over the four-volume gives the Euler Characteristic. As an action linearly depending on $\G$ does not provide any contribution to the equations of motion in four dimensions (or less), a function of the Gauss-Bonnet term $f(\G)$ turns out to be non-trivial even in 3+1 dimensions. Due to the presence of the topological term, $f(\G)$  gravity yields a reduced dynamics with respect to other modified theories. Moreover, it fits cosmological observations even in four dimensions \cite{Bajardi:2020osh} and provides generalizations of the Newtonian potential in the weak field limit \cite{Bajardi:2019zzs}. The variation of the action
\begin{equation}
S = \frac{1}{2 \kappa}\int \sqrt{-g} f(\G) \, d^4x +\mathcal{S}^{(m)},
\label{f(G)}
\end{equation}
with respect to the metric tensor yields
\begin{equation}
\begin{split}
&2R D_\mu D_\nu f_\G(\G) - 2g_{\mu \nu} R \Box f_\G(\G) - 4R^\lambda_\mu D_\lambda D_\nu f_\G(\G) + 4R_{\mu \nu} \Box f_\G(\G) 
\\
&+ 4 g_{\mu \nu} R^{p \sigma} D_p D_\sigma f_\G(\G) + 4 R_{\mu \nu p \sigma} D^p D^\sigma f_\G(\G) + \frac{1}{2} g_{\mu \nu}[f(\G) - \G f_\G(\G)] = \kappa T_{\mu \nu} \;.
\end{split}
\label{field equations}
\end{equation}
The Gauss--Bonnet invariant naturally arises in gauge theories of gravity such as the Lovelock \cite{Lovelock:1971yv, Lovelock:1972vz} or the Chern-Simons gravity \cite{Zanelli:2005sa, Achucarro:1987vz}. Often the Ricci scalar is additively included into the action \eqref{f(G)}, so that GR can be safely recovered as soon as the function $f(\G)$ vanishes or is negligible. However, as pointed out in \cite{Bajardi:2020osh}, the action \eqref{f(G)} yields  physically relevant results provided by the  standard cosmology even without imposing the GR limit as a requirement. In other words, GR dynamics can be recovered also from a pure $f(\G)$ action.

Another class of extensions takes into account  higher-order curvature invariants into the action. It is introduced to consider higher loop corrections in QFT formulated in curved spacetimes \cite{Jurgen}. The action is 
\begin{equation}
S = \int \sqrt{-g} \, f(R, \Box R, \Box^2 R...\Box^k R) \, d^4 x,
\label{HOACT}
\end{equation}
the variational principle provides the following field equations \cite{Capozziello:2011et}
\begin{eqnarray}
\nonumber
&& G^{\mu \nu} = \frac{1}{{\cal{M}}}\left\{\frac{1}{2} g^{\mu\nu} (F - {\cal{M}} R)+ (g^{\mu \lambda} g^{\nu \sigma} - g^{\mu \nu} g^{\lambda \sigma}){\cal{M}}_{; \lambda ;\sigma}  \right. 
\\
&&\nonumber
+ \frac{1}{2} \sum_{i = 1}^k \sum_{j = 1}^i (g^{\mu \nu} g^{\lambda \sigma} + g^{\mu \lambda} g^{\nu \sigma})(\Box^{j-1} R)_{; \sigma} \left(\Box^{i-j} \frac{\partial F}{\partial \Box^i R} \right)_{; \lambda}   
\\
&&\left. - g^{\mu \nu} g^{\lambda \sigma} \left[(\Box^{j-1} R)_{; \sigma} \;  \left(\Box^{i-j} \frac{\partial F}{\partial \Box^i R} \right) \right]_{; \lambda} \right\},
\label{eq di campo per ordine sup al quarto}
\end{eqnarray}
being $\cal{M}$ defined as
\begin{equation}
{\cal{M}} \equiv \sum_{j = 0}^k \left(\Box^{j} \frac{\partial F}{\partial \Box^j R} \right).
\end{equation}

It can be showed that, under conformal transformations of the form $g_{\mu \nu} \to \tilde{g}_{\mu \nu} = e^{2 \omega} g_{\mu \nu}$, the $(2k+4)^{th}$ order action \eqref{HOACT} is formally equivalent to a $(2k+2)^{th}$-order theory where geometry is non-minimally coupled to scalar field $\phi$ \cite{Jurgen, Uzan:1999ch, Bartolo:1999sq, DeFelice:2011hq}. 

The most general second-order scalar-tensor theory was proposed by Horndeski in \cite{Horndeski:1974wa} and the corresponding action reads \cite{Deffayet:2009mn, Deffayet:2009wt, Capozziello:2018gms, Copeland:2018yuh} 
\begin{equation}
\label{ETG:10}
\mathcal{S}_{Horndeski}=\sum_{i=2}^5 \int\!d^4x \sqrt{-g}\,\mathcal{L}_i\,,
\end{equation}
with the definitions
\begin{align*}
&\mathcal{L}_2\equiv G_2(\phi,X) \,,\\
&\mathcal{L}_3\equiv -G_3(\phi,X)\,\square\phi \,,\\
&\mathcal{L}_4\equiv G_4(\phi,X)R+G_{4X}\,\bigl[(\square\phi)^2-(D_{\mu}D_{\nu}\phi)^2\bigr]\,, \\
&\mathcal{L}_5\equiv G_5(\phi,X)\,G_{\mu\nu}D^{\mu}D^{\nu}\phi-\frac{1}{6}\,G_{5X}\,\bigl[(\square\phi)^3%
                          -3\,\square\phi\,(D_{\mu}D_{\nu}\phi)^2+2\,(D_{\mu}D_{\nu}\phi)^3\bigr] \,.
\end{align*}
The functions $G_i(\phi,X)$ are arbitrary functions of the scalar field and of the kinetic term: %
$X\equiv\frac{1}{2}g^{\mu\nu}D_{\mu}\phi D_{\nu}\phi$\,, $\displaystyle{G_{iX}\equiv\frac{\partial G_i}{\partial X}}$. 

As an example, by setting $G_2=\frac{\omega}{\phi}X,\,G_3=0,\,G_4=\phi,G_5 = 0$ the Brans-Dicke theory is recovered, while for $G_2=X-V(\phi),\,G_3=0,\,G_4=F(\phi),\,G_5 = 0$, one recovers the standard scalar-tensor theory of gravity with coupling function $F(\phi)$ and potential $V(\phi)$. 
The above examples represent most of extended theories of gravity where, with this terminology, we intend  theories which, in some way,  extend GR by improving the Hilbert-Einstein action by geometric invariants or scalar fields. In all these cases, dynamics is curvature based and the metric $g_{\mu\nu}$ represents the gravitational field.

Besides the standard description, GR can be described by two other equivalent formalisms, which yield the same dynamics as Einstein gravity. One of them is the above mentioned TEGR, where the gravitational interaction is represented by means of torsion instead of curvature. The other is the so called \emph{Symetric Teleparallel Equivalent to General Relativity} (STEGR), where the spacetime is ruled by the so called \emph{non-metricity}. 

In purely metric theories, such as GR, the causal structure and the geodesic structure are uniquely determined by the metric $g_{\mu\nu}$ only. In particular, the Levi-Civita connection is based on two main assumptions:
\begin{itemize}
\item  it is metric-compatible, namely requires $D_{\!\!\rho}\,g_{\mu\nu}=0$\,;
\item  it is symmetric, \emph{i.e.} $\Gamma^{\,\rho}_{\,\,\,\mu\nu}= \Gamma^{\,\rho}_{\,\,\,\nu\mu}$\,.
\end{itemize}
Once that such assumptions are relaxed, it is possible to define a \emph{Non-Metricity Tensor} $Q_{\rho\mu\nu}\equiv {D}_{\!\!\rho}\,g_{\mu\nu} \ne0$ and a \emph{Torsion Tensor} $T^\alpha_{\,\,\, \mu \nu} = 2 \Gamma^\alpha_{[\mu \nu]}$, which can be included in the most general connection 
\begin{equation}
\label{ETG:20}
\Gamma^{\rho}_{\,\,\,\,\mu\nu}=\overset{\!\!\!\!\!\circ}{\Gamma^{\,\rho}}_{\mu\nu}+K^{\rho}_{\,\,\,\,\mu\nu}+L^{\rho}_{\,\,\,\,\mu\nu}.
\end{equation}
In the above equation $\overset{\!\!\!\!\!\circ}{\Gamma^{\,\rho}}_{\mu\nu}$ is the Levi-Civita connection, $K^{\rho}_{\,\,\,\mu\nu}$ is the so-called \emph{contorsion tensor}
\begin{equation}
K^{\rho}_{\,\,\,\,\mu\nu}\equiv\frac{1}{2}g^{\rho\lambda}\bigl(T_{\mu\lambda\nu}+T_{\nu\lambda\mu}+T_{\lambda\mu\nu}\bigr)%
                                      =-K^{\rho}_{\,\,\,\,\nu\mu} \,,
\end{equation}
and $L^{\rho}_{\,\,\,\mu\nu}$ is the so-called \emph{disformation tensor}
\begin{equation}
L^{\rho}_{\,\,\,\,\mu\nu}\equiv\frac{1}{2}g^{\rho\lambda}\bigl(-Q_{\mu \nu \lambda}-Q_{\nu \mu \lambda} + Q_{\lambda\mu\nu}\bigr)=%
                                               L^{\rho}_{\,\,\,\,\nu\mu}\,.
\end{equation}
As the Levi-Civita connection is related to curvature, the contorsion and disformation tensor can be addressed to torsion and non-metricity, respectively. In the context of GR, both $L^{\rho}_{\,\,\,\,\mu\nu}$ and $K^{\rho}_{\,\,\,\,\mu\nu}$ vanish. Specifically, we have:
\begin{equation}
 \begin{array}{l} \mbox{GR} \rightarrow L^{\rho}{ }_{\mu \nu}=K^{\rho}{ }_{\mu \nu} =0\,, \\ \mbox{TEGR} \rightarrow \overset{\circ}{\Gamma}\,^{\rho}{ }_{\mu \nu}, \,\, K\,^{\rho}{ }_{\mu \nu} \neq 0, \quad L^{\rho}{ }_{\mu \nu}=0\, , \\ \mbox{STEGR} \rightarrow \overset{\circ}{\Gamma}\,^{\rho}{ }_{\mu \nu},  L^{\rho}{ }_{\mu \nu}\neq 0, \quad K^{\rho}{ }_{\mu \nu}=0 \,.\end{array}
\end{equation}
From a physical point of view, the presence of curvature in the spacetime means that the final orientation of a given vector parallel transported along a closed path is different than the initial one. The presence of torsion leads in turn to a shift of the vector after performing a closed path. Finally, when the spacetime is labeled by non-metricity, the length of the vector changes during the path.
\begin{center}
\centering
\includegraphics[width=.80\textwidth]{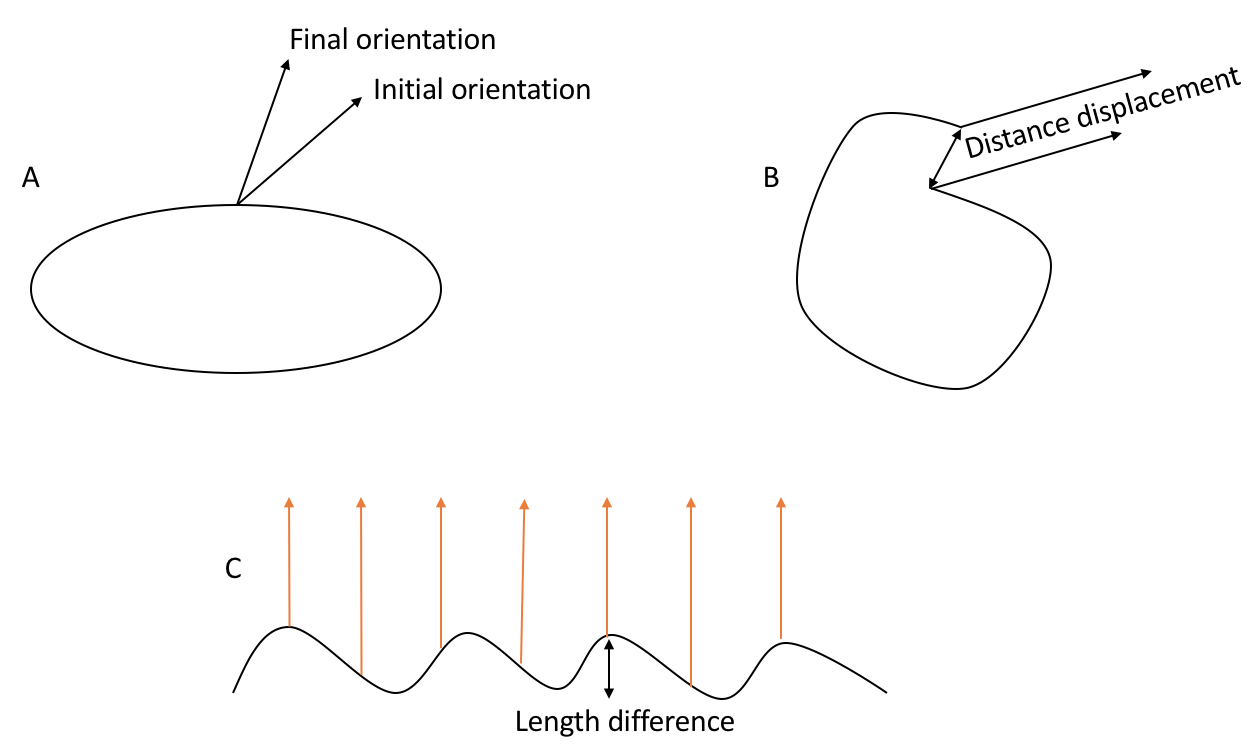}
\\ Figure 1: \emph{Difference between initial and final position of a vector after a parallel transport in different spacetimes. Fig. A describes a spacetime with curvature and without torsion and non-metricity;  Fig. B relies to a spacetime only labeled by torsion; in Fig. C the spacetime is only characterized by the presence of non-metricity.}
\end{center}
In STEGR it is possible to choose the so called \emph{coincident gauge}, where the total connection is trivialized, namely $\Gamma^\rho_{\,\,\, \mu \nu} = 0$. In this gauge, by defining the \emph{Superpotential}, the \emph{Torsion Scalar} and the \emph{Non-Metricity Scalar} respectively as:
\begin{equation}
S^{p \mu \nu} \equiv K^{\mu \nu p}-g^{p \nu} T_{\,\,\,\,\,\,\, \sigma}^{\sigma \mu}+g^{p \mu} T_{\,\,\,\,\,\,\, \sigma}^{\sigma \nu},
\end{equation}
\begin{equation}
T \equiv T_{p \mu \nu} S^{p \mu \nu},
\end{equation}
\begin{equation}
Q \equiv  -\frac{1}{4} Q_{\alpha \mu \nu} \left[- 2 L^{\alpha \mu \nu}+  g^{\mu \nu} \left(Q^\alpha - \tilde{Q}^\alpha \right) - \frac{1}{2} \left(g^{\alpha \mu} Q^\nu + g^{\alpha \nu} Q^\mu  \right)\right],
\end{equation}
with
\begin{eqnarray}
&& Q_\mu \equiv Q_{\mu \,\,\,\, \lambda}^{\,\,\,\lambda},
\\
&& \tilde{Q}_{\mu}\equiv Q_{\alpha \mu}^{\, \, \, \, \, \, \alpha},
\end{eqnarray}

the GR, TEGR and STEGR actions 
\begin{eqnarray}
&&\mathcal{S}_{GR}\equiv \frac{1}{2 \kappa}  \int d^4x\,\sqrt{-g}\,\overset{\circ}{R}+\mathcal{S}^{(m)}\,,\\
&&\mathcal{S}_{TEGR}\equiv \frac{1}{2 \kappa} \int d^4x\,\sqrt{-g}\,T+\mathcal{S}^{(m)}\,\label{TGACT},\\ 
&&\mathcal{S}_{STEGR}\equiv \frac{1}{2 \kappa} \int d^4x\,\sqrt{-g}\,Q+\mathcal{S}^{(m)}\,,
\end{eqnarray}
only differ for a boundary term. Here $\overset{\circ}{R}$ is the scalar curvature written in terms of the Levi-Civita connection. From the above result it follows that the three theories are dynamically equivalent and physically indistinguishable at the level of equations. For this reason, they are usually thought as parts of the same structure, the so called \emph{Geometric Trinity of Gravity} \cite{Koivisto,Jackson}. While STEGR cannot be recast in terms of a gauge theory in a locally flat spacetime, GR and TEGR can be treated as gauge theories of the local Lorentz group and the translation group, respectively. The gauge-invariance of TEGR, can be investigated by defining the \emph{tetrad fields} $h^a_\mu$ as 
\begin{equation}
g_{\mu \nu} = h^a_\mu h^b_\nu \eta_{ab},
\end{equation}
where Latin indexes label the flat spacetime and Greek indexes the curved spacetime. Tetrad fields are a mathematical tool capable of linking the curved spacetime with the locally flat spacetime point by point. It is possible to show (see \cite{Aldrovandi:2015wfa, Aldrovandi:2013wha} for further details) that in the reference frame in which the spin connection vanishes, TEGR connection can be written in terms of tetrad fields as
\begin{equation}
\Gamma_{\,\,\, \mu \nu}^{\rho}=h_{a}^{\rho} \partial_{\mu} h_{\nu}^{a},
\end{equation}
and is called \emph{Weitzenb$\ddot{o}$ck connection}. In this way, the explicit expression of the torsion tensor is
\begin{equation}
T_{\,\,\, \mu \nu}^{\rho}= h_{a}^{\rho} \partial_{\mu} h_{\nu}^{a}-h_{a}^{\rho} \partial_{\nu} h_{\mu}^{a},
\label{Torsion}
\end{equation}
so that the relation between torsion scalar $T$ and scalar curvature $\overset{\circ}{R}$ can be written as
\begin{equation}
\overset{\circ}{R} -\frac{2}{h} \partial_{\mu}\left(h T_{\,\,\,\,\,\, \nu}^{\nu \mu}\right)=-T,
\label{RBT}
\end{equation}
with $h$ being the determinant of the tetrad fields. In particular, by means of the definition $B \equiv \displaystyle \frac{2}{h} \partial_{\mu}\left(h T_{\,\,\,\,\,\, \nu}^{\nu \mu}\right)$, the relation \eqref{RBT} takes the simple form
\begin{equation}
\overset{\circ}{R} = - T + B.
\label{R = -T+B}
\end{equation}
The field equations coming from Eq. \eqref{TGACT} can be found by varying the action with respect to the tetrads. They read:
\begin{equation}
\frac{4}{h} \partial_{\mu}\left(h S_{a}^{\,\,\, \mu \beta}\right)-4 T_{\,\,\, \mu a}^{\sigma} S_{\sigma}^{\,\,\, \beta \mu}-T h_{a}^{\beta} = 0,
\label{TEGRFE}
\end{equation}
and are formally equivalent to Einstein field equations. It is worth noticing that the theory is invariant under the local translation group by construction. In this way, the torsion scalar accounts for the gauge field related to the invariance under translations. As a consequence, the Equivalence Principle is not assumed as a requirement \emph{a priori}. This means that some aspects of the theory might hold even beyond the Planck scale, where the  Equivalence Principle and Lorentz invariance could be violated.

In the large-scale regime, TEGR and STEGR suffer same shortcomings as GR. For this reason, in analogy to GR modifications, extended teleparallel actions started being considered as viable modifications of TEGR. The teleparallel equivalent of $f(R)$ gravity can be addressed by introducing into the action a function of the torsion scalar, that is:
\begin{equation}
S = \int h \, f(T) \, d^4x.
\end{equation} 
By varying the action with respect to the tetrad fields, one gets the following field equations \cite{Cai:2015emx}
\begin{equation}
\frac{1}{h} \partial_\mu(h \; h^p_a S_p^{\;\; \mu \nu}) f_T(T) - h^\lambda_a T^p_{\;\; \mu \lambda} S_p^{\;\; \nu \mu} f_T(T)+ h^p_a S_p^{\;\; \mu \nu}(\partial_\mu T) f_{TT}(T) + \frac{1}{4} h^\nu_a f(T) = 0,
\end{equation}
which reduce to those in Eq. \eqref{TEGRFE} when $f(T) = T$. 

Several teleparallel modified theories of gravity have been considered so far in the literature, as well as the coupling between torsion and scalar fields \cite{Yang:2010ji, Hohmann:2018rwf} or the Teleparallel equivalent of Gauss-Bonnet theory \cite{Kofinas:2014owa, Kofinas:2014aka, Capozziello:2016eaz}. In this paper (Sec. \ref{BoxT}), we consider a non-local action including the torsion scalar and the boundary term and, after selecting viable models by Noether symmetry approach, we find out exact cosmological solutions. 

It is worth noticing that, although TEGR and GR are dynamically equivalent, the above mentioned extensions of TEGR do not provide the same dynamics as GR extensions. As an example, the $f(R)$ theory leads to fourth-order field equations, while the field equations of $f(T)$ gravity are of the second-order. Moreover, as well as standard TEGR, $f(T)$ gravity is invariant under infinitesimal translations, instead of Lorentz transformations (like $f(R)$ gravity). This is due to the fact that curvature can be recast as the gauge field related to the Lorentz group, while torsion is the gauge field coming from the translation group.

However, though several modified theories of gravity (involving curvature, torsion or non-metricity) are power-counting renormalizable, in most cases they are not unitary as other interactions, and cannot be treated under the Yang-Mills standard. So far a comprehensive and self-consistent unitary theory of gravity, also capable of solving the cosmological and astrophysical problems, is still missing. In this regard, non-local theories of gravity can represent  valuable candidates towards the construction of a self-consistent theory of gravity, also fitting the cosmological and astrophysical observations \cite{Modesto:2017hzl, Tomboulis:2015esa}. A further discussion is relied to the next section, where the main properties of non-local theories of gravity are summarized.

\subsection{Non-Locality in  Theories of Gravity}
Let us briefly introduce the principal features of non-local theories of gravity.  The main difference with respect to the corresponding local  theories, is that the effective action contains non-local operators of different form, which aim to merge the gravitational interaction with the quantum formalism.

Depending on the type of non-locality, non-local  theories of gravity can be classified in two main classes: Infinite Derivative Theories of Gravity (IDGs) and Integral Kernel Theories of Gravity (IKGs). 

The former  involve  analytic transcendental functions of the covariant d'Alembert operator $\square$. An example is given by the model proposed in Ref. \cite{Modesto:2013ioa}, which provides a solution for classical black hole and Big Bang singularities \cite{Modesto:2011kw, Briscese:2012ys}. 

On the other hand, IKGs mainly adopt the inverse operator $\square^{-1}$. They were firstly considered in \cite{Deser:2007jk},  where it is 
  shown that the application of the non-local operator $\square^{-1}$ to the scalar curvature $R$, gives rise to the  late-time cosmic expansion of the universe without invoking any Dark Energy contribution. In  view of merging gravity with the other fundamental interactions, IDGs provide renormalizable and unitary quantum gravity theories \cite{Biswas:2011ar}, while IKGs deal with infrared (IR) quantum corrections coming from the formulation of QFT in curved spacetime \cite{Barvinsky:2014lja}. Despite these nice  features, no local and non-local theory capable of solving all the large-scale structure issues and fitting all the today available observations occurred so far.
\subsubsection{Infinite Derivative Theories of Gravity}
In order to introduce the main aspects of IDGs, let us consider the following infinite-derivatives Lorentz-invariant action:
\begin{equation}
S=\frac{1}{2} \int d^{4} x d^{4} y \phi(x) \mathcal{K}(x-y) \phi(y)-\int d^{4} x V(\phi),
\end{equation} 
where $\phi$ is a scalar field depending on the coordinates and $\mathcal{K}(x-y)$, an operator depending on the distance $x-y$ through a generic function of the d'Alembert operator $\Box$, as
\begin{equation}
\mathcal{K}(x-y)=F(\square) \delta^{(4)}(x-y).
\end{equation}  
A straightforward factorization of $F(\Box)$, yields
\begin{equation}
F(\square)=e^{-\gamma(\square)} \prod_{i=1}^{N}\left(\square-m_{i}^{2}\right),
\end{equation}
with $\gamma(\square)$ being an entire function. By means of a Fourier transformation, it is possible to show that  ghosts appear when $N >1$. For this reason, hereafter we focus on the $N=1$ choice, where unitarity is preserved. The most general action made of functions of the d'Alembert operator, which is ghost-free and quadratic in the curvature, must contain infinite covariant derivatives. It reads \cite{Biswas:2005qr, Biswas:2011ar, Biswas:2016etb, Biswas:2016egy}: 
\begin{equation}
\mathcal{S}=\frac{1}{2 \kappa} \int \!d^4x\, \sqrt{-g}\,\Bigl[R+\alpha\Bigl(RF_1(\square_s)R+R_{\mu\nu}F_2(\square_s)R^{\mu\nu}+R_{\mu\nu\rho\sigma}F_3(\square_s)R^{\mu\nu\rho\sigma}\Bigr)\Bigr]   \,,
\label{action0}
\end{equation}
where $F_i(\square_s)$ is a transcendental  analytic functions of $\square_s\equiv\square/M^2_s$, with $M_s$ being a mass/lenght scale introduced to make the d'Alembert operator dimensionless. Ghosts are prevented by the nature of the  function, which manifests no pole on the whole complex plane. 

In Ref. \cite{Modesto:2013ioa} a subcase of Eq. \eqref{action0} is proposed, which provides GR at the zero order:
\begin{equation}
\mathcal{S}=\frac{1}{2 \kappa} \int \!d^4x\, \sqrt{-g}\,\biggl(R-G_{\mu\nu}\frac{e^{H(-\square_s)}-1}{\square}R^{\mu\nu}\biggr) \,,
\end{equation}
where $H(-\square_s)$ is an  analytic function of $\square_s$. The field equations, up to the order $O(R^2)$, reads
\begin{equation}
G_{\mu\nu}+O(R^2)=\kappa  \, e^{-H(-\square_s)}\,T_{\mu\nu}^{\,(m)}.
\label{feMOD}
\end{equation}
Notice that, at the lowest order of the Taylor expansion, the field equations are equivalent to those of GR. In spherical symmetry, Eq. \eqref{feMOD} provides regular black holes without singularities \cite{Modesto:2011kw}, while in an homogeneous and isotropic universe, it admits bouncing cosmological solutions \cite{Briscese:2012ys}. The mechanism aimed at avoiding classical singularities is discussed in details in Refs. \cite{Buoninfante:2018xiw, Buoninfante:2018mre}. It basically consists in a non-local smearing of the point-like source of the Schwarzschild background, which automatically implies that the metric can no longer be considered as a vacuum solution. In the non-local region, where $r<2/M_s$, the effects of non-locality start being relevant, while, as soon as the radius start to increase,  the solution  approaches the Schwarzschild one. As mentioned above, the theory does not contain singularities for any value of $r$, there including $r=0$. 

Another example is provided in Ref. \cite{Briscese:2013lna}. This theory turns out to be unitary and renormalizable and represents the maximal UV-completion for the Starobinsky gravity. The corresponding action reads as:
\begin{equation}
\mathcal{S}=\frac{1}{2 \kappa}\int \!d^4x\, \sqrt{-g}\,\biggl[R-G_{\mu\nu}\frac{V_2^{-1}-1}{\square}R^{\mu\nu}+\frac{1}{2}R\,\frac{V_0^{-1}-V_{2}^{-1}}{\square}\,R\biggr] \,,
\end{equation}
where
\begin{equation*}
V_2^{-1}\equiv e^{H_2(-\square_s)}p^{(n_2)}(-\square_s) \,,\quad V_0^{-1}-V_{2}^{-1}\equiv\frac{1}{3}\Bigl[e^{H_0(-\square_s)}(1+\square_s)-e^{H_2(-\square_s)}\Bigr] \,,
\end{equation*}
with $H_i$ and $p^{(n_2)}$ being  analytic functions. The cosmological applications are extremely interesting because the related effective lengths can be useful for large scale structure as well as  IR behavior.

\subsubsection{Integral Kernel Theories of Gravity}
The other family of non-local theories of gravity is represented by IDGs. They bring UV quantum corrections by the expansion around $s=0$ of a Schwinger proper time integral \cite{Birrell:book}. On the other hand, IR corrections need an expansion around $s\to\infty$, which may represent a problem for several reasons. First, the Schwinger proper time integral can be settled only when masses of matter fields are greater than the potential; second, in the massless limit, the proper time integration starts increasing for late times, up to infinity. These issues come directly from the perturbative approach used to calculate the Schwinger proper time integral, thus a non-perturbative technique is needed in order to obtain both UV ($s=0$) and IR ($s\to\infty$) corrections. The quantum effective action coming from such non-perturbative technique reads \cite{Barvinsky:2014lja} 
\begin{equation}
W_0=-\int\!d^4x \sqrt{-g}\,\Bigl[V(x)+V(x)(\square-V)^{-1}V(x)\Bigr]+\frac{1}{6}\,\Sigma \,,
\label{actiono0}
\end{equation}
where $V(x)$ is a generic potential and $\Sigma$ is a surface term defined through the inverse of d'Alembert operator as \cite{Barvinsky:2014lja}
\begin{equation}
\begin{split}
\Sigma=\frac{1}{2 \kappa} \int\!d^4x \sqrt{-g}\,\Bigl\{R&-R_{\mu\nu}\,\square^{-1}G^{\mu\nu}+ %
                                           2^{-1} R\,\bigl(\square^{-1}R^{\mu\nu}\bigr) \square^{-1} R_{\mu\nu}  \\
             &-R^{\mu\nu}\bigl(\square^{-1} R_{\mu\nu}\bigr) \square^{-1}R+ %
                \bigl(\square^{-1} R^{\alpha\beta}\bigr) \bigl(D_{\!\alpha}\,\square^{-1}R\bigr) D_{\!\beta}\,\square^{-1}R \\
             &-2\,\bigl(D^{\mu}\,\square^{-1} R^{\nu\alpha}\bigr) \bigl(D_{\!\nu}\,\square^{-1} R_{\mu\alpha}\bigr) \square^{-1}R \\
             &-2\,\bigl(\square^{-1} R^{\mu\nu}\bigr) \bigl(D_{\!\mu}\,\square^{-1}R^{\alpha\beta}\bigr) %
                 D_{\!\nu}\,\square^{-1} R_{\alpha\beta}+O\bigl[R_{\mu \nu}^{\,\,4}\bigr]\Bigr\} \,.
\end{split}
\end{equation} 

In this case, the non-locality is due to the integral operator $\square^{-1}$, which is  responsible for the quantum corrections. However, the action \eqref{actiono0} provides non-linear higher-order field equations, which may represent an obstacle towards the search for exact solutions. For this reasons, several subcases have been treated in the literature, among which the straightforward correction to GR \cite{Deser:2007jk}
\begin{equation}
\label{NL:8}
\mathcal{S}=\frac{1}{2 \kappa} \int\!d^4x \sqrt{-g}\,R\,\Bigl[1+F\Bigl(\square^{-1}R\Bigr)\Bigr] +\mathcal{S}^{(m)}\,,
\end{equation}
with $F\Bigl(\square^{-1}R\Bigr)$ being an arbitrary function of the non-local term $\square^{-1}R$. The variation of the above action with respect to the metric tensor yields:
\begin{equation}
G_{\mu\nu}+\Delta G_{\mu\nu}=\kappa  T_{\mu\nu}^{\,(m)} \,,
\end{equation}
with the definitions
\begin{eqnarray}
&& \Delta G_{\mu\nu}=\,\Bigl(G_{\mu\nu}+g_{\mu\nu}\,\square-D_{\mu}D_{\nu}\Bigr)%
                                \biggl\{F+\square^{-1}\Bigl[R\,F'\Bigr]\biggr\}  \nonumber 
                                \\
                          && \qquad \qquad   +\biggl[\delta_{\mu}^{\,\,(\rho}\,\delta_{\nu}^{\,\,\,\sigma)}-\frac{1}{2}\,g_{\mu\nu}g^{\rho\sigma}\biggr]%
                                \partial_{\rho}\Bigl(\square^{-1}R\Bigr)\,\partial_{\sigma}\biggl(\square^{-1}\Bigl[R\,F'\Bigr]\biggr)   ,               
\\
&&                          F\equiv F\Bigl(\square^{-1}R\Bigr), \qquad F'\equiv\frac{\partial F}{\partial\bigl(\square^{-1}R\bigr)}.
\end{eqnarray}
It is possible to show that the presence of the operator $\square^{-1}$ naturally explains late-time cosmic acceleration of the universe. To this purpose, let $t_0\sim10^{10} y$ be the present time, $t_{eq}\sim10^5 y$ the time of the equivalence and assume a matter dominated era between $t_0$ and $t_{eq}$. In a spatially-flat homogeneous and isotropic universe, simple computations show that the contribution of non-local causal effects, acting within the interval considered, are of the order of
\begin{equation}
\bigl(\square^{-1}R\bigr)(t_0)\,\,\sim 14.0\,,
\end{equation}
which is the large number required by the current cosmic acceleration to avoid the fine tuning of parameters. 

In next sections we will focus on higher-order IKGs, where non-locality is induced by the operator $\Box^{-1}$. More precisely, they are generalizations of the IKGs considered \emph{e.g.} in \cite{Nojiri:2007uq, Deser:2007jk, Bahamonde:2017sdo, Jhingan:2008ym} . Even though we will not discuss in details the UV quantum corrections of these theories, we expect that the presence of non-local operators is somehow useful in order to recover renormalizability and unitarity. In view of obtaining exact cosmological solutions, we reduce the dynamics by the search of Noether symmetries, a useful approach developed to select viable models, whose main aspects are outlined in \ref{noeth}.

\section{Curvature-based Non-Local Gravity: The Case $F(R, \Box^{-1} R)$} \label{BoxR}
As an example of the above considerations, let us  study the  metric  non-local IKG gravity  given by the action
\begin{equation}
\label{NL:22}
\mathcal{S}=\frac{1}{2 \kappa} \int\!d^4x \sqrt{-g}\,F\bigl(R,\square^{-1}R\bigr)\,.
\end{equation}
It is a straightforward  extension of  both $f(R)$ gravity and  action \eqref{NL:8}. In order to construct a point-like Lagrangian useful for cosmological considerations, we define an  auxiliary \textit{local} scalar field $\phi$ defined as
\begin{equation}
\label{NL:23}
\phi\equiv\square^{-1}R, \quad \textup{so that} \quad  R\equiv\square\phi \,.
\end{equation}
By means of this definition, the theory reduces to a class of higher-order scalar-tensor models with action
\begin{equation}
\label{NL:24}
S= \int d^4x\,\sqrt{-g}\,\,F(R,\phi),
\end{equation}
where the constant $1/2 \kappa$ is included into the function $F$. The action in Eq. \eqref{NL:24} represents a generalization of that considered in Ref.\cite{Nojiri:2007uq}, 
where the authors start from the Deser-Woodard action to find exact cosmological solutions. From this point of view, the advantage of  Noether Symmetry Approach is to select the action by a physical criterion, among several possible choices. Though the Deser-Woodard action is contained in Eq. \eqref{NL:24}, several other models can be selected by  symmetries. In any case, the related conserved quantities allow to reduce the dynamics and to find out analytic solutions. By using the cosmological expression of $R$ and $\Box R$ in a Friedmann-Lema\^itre-Robertson-Walker (FLRW) spacetime, the action can be written as:
\begin{equation}
\label{NL:27}
S= \frac{\pi^2}{\kappa} \int\!dt\,a^3\biggr\{F(R,\phi)-\lambda_1(R-\ddot{\phi}-3H\dot{\phi})-\lambda_2 \biggl[R+6\biggl(\frac{\ddot{a}}{a}+\Bigl(\frac{\dot{a}}{a}\Bigr)^{\!2}\biggr) \biggr]\biggr\},
\end{equation}
where $\lambda_1$ and $\lambda_2$ are  Lagrange multipliers. The relation between $\lambda_1$ and $\lambda_2$ can be found by varying the action with respect to $R$: it reads
\begin{equation}
\lambda_2=\frac{\partial F(R,\phi)}{\partial R}-\lambda_1.
\end{equation}
Following the same procedure \emph{e.g.} as \cite{Nojiri:2007uq, Elizalde:2018qbm}, the constant $\lambda_1$ can be promoted to a time-depending scalar field by setting $\lambda_1\equiv \epsilon(t)$. In this way, Eq. \eqref{NL:27} can be recast as:
\begin{equation}
\label{NL:30}
S=\frac{\pi^2}{\kappa}  \!\int\!dt\,a^3 \biggr\{F(R,\phi)-\epsilon(R-\ddot{\phi}-3H\dot{\phi})-\biggl(\frac{\partial F(R,\phi)}{\partial R}-\epsilon\biggr)\biggl[R+6\biggl(\frac{\ddot{a}}{a}+\Bigl(\frac{\dot{a}}{a}\Bigr)^{\!2}\biggr) \biggl]\biggr\}.
\end{equation}
As standard when dealing with Lagrange multipliers method, we consider $\phi$ and $R$ as independent fields. Furthermore, the variation of the action with respect to the scalar fields $\phi$ and $\epsilon$, provide the Klein-Gordon equations
\begin{equation}
\square \epsilon(t)=F_{\phi}(R, \phi), \qquad \Box \phi = R,
\label{KGEQ}
\end{equation}
respectively. Integrating out the second derivatives of the scalar fields, the point-like Lagrangian written in the configuration space ${\cal Q}\equiv \{a(t), R(t), \phi(t), \epsilon(t)\}$, reads as:
\begin{equation}
\Lagr =\,\,a^3F-a^3\dot{\phi}\dot{\epsilon}-a^3 R\,\partial_R F+6a\dot{a}^2\partial_R F\,
    -6a\dot{a}^2\epsilon+6a^2\dot{a}\dot{R}\,\partial_{RR}F+6a^2\dot{a}\dot{\phi}\,\partial_{R\phi}F-6a^2\dot{a}\dot{\epsilon}.
    \label{lagrf(R)}
\end{equation}
Notice that not all the Euler-Lagrange equations provide  contributions to the dynamics. The equations with respect to $\epsilon$ and $\phi$ give back the Klein-Gordon equations \eqref{KGEQ}. 
The equation with respect to the scalar curvature provides the cosmological expression of $R$ by construction. The Euler-Lagrange equation with respect to the scale factor together with the energy condition are the only dynamical equations of motion, by means of which analytic solutions can be found. They correspond to the "0,0" and "1,1" components of the field equations. 

In the related configuration space (the minisuperspace), the  symmetry generator is
\begin{eqnarray}
\label{eqn:X[1]}
\mathcal{X} &=& \xi(t,a,\phi,R,\epsilon) \frac{\partial}{\partial t} + \alpha(t,a,R,\phi,\epsilon) \frac{\partial}{\partial a}+\beta(t,a,R,\phi,\epsilon) \frac{\partial}{\partial R}  \nonumber
\\
&+&\gamma(t,a,R,\phi,\epsilon)\frac{\partial}{\partial\phi}+\delta(t,a,R,\phi,\epsilon) \frac{\partial}{\partial \epsilon}.
\end{eqnarray}
                                                        
Therefore, by applying the Noether symmetry existence condition \eqref{noethid} to the Lagrangian \eqref{lagrf(R)}, we get a system of 28 differential equations. Only six of them are linearly independent, namely
\begin{eqnarray}
&&\begin{split}
     \alpha\,\partial_RF&-\alpha \epsilon+a\beta\,\partial_{RR}F+a\gamma\,\partial_{R\phi}F-a\delta+2a\,\partial_RF\,\partial_a\alpha%
     -2a\epsilon\,\partial_a\alpha \\&+a^2\partial_{RR}F\,\partial_a\beta+a^2\partial_{R\phi}F\,\partial_a\gamma%
     -a^2\partial_a\delta-a\,\partial_RF\,\partial_t\xi+a\epsilon\,\partial_t\xi=0, \qquad\qquad                   \nonumber              
     \end{split}                                                                                                                                                                             \\
&&\begin{split} 
     2\alpha\,\partial_{RR}F+a\beta\,\partial_{RRR}F+a\gamma\,\partial_{RR\phi}F&+a\,\partial_a\alpha\,\partial_{RR}F%
      \\&+a\,\partial_R\beta\,\partial_{RR}F-a\,\partial_t\xi\,\partial_{RR}F=0 ,                            \nonumber 
     \end{split}                                                                                                                                                             \\
&&\begin{split}
     12\alpha\,\partial_{R\phi}F&+6a\beta\,\partial_{RR\phi}F+6a\gamma\,\partial_{R\phi\phi}F%
     +6a\,\partial_a\alpha\,\partial_{R\phi}F \\&+6a\,\partial_{\phi}\beta\,\partial_{RR}F+6a\,\partial_{\phi}\gamma\,\partial_{R\phi}F%
     -a^2\partial_a\delta-6a\,\partial_{R\phi}F\,\partial_t\xi=0     ,                                                    \nonumber                   
     \end{split}                                                                                                                                                                               \\
&&-12\alpha-6a\,\partial_a\alpha+ 6a\,\partial_{\epsilon}\beta\,\partial_{RR}F-a^2\partial_a\gamma \nonumber
     -6a\,\partial_{\epsilon}\delta+6a\,\partial_t\xi=0,                                                                                           \\
&&-3\alpha-a\,\partial_{\phi}\gamma- a\,\partial_{\epsilon}\delta+a\,\partial_t\xi=0,           \nonumber  \\
&&\begin{split}
     3\alpha F-3\alpha R\,\partial_RF-aR\,\beta\,\partial_{RR}F+a\gamma\,\partial_{\phi}F&-aR\,\gamma\,\partial_{R\phi}F %
     \\&+ a F\,\partial_t\xi-aR\,\partial_RF\,\partial_t\xi=0 \nonumber.                              
     \end{split}
      \\ \nonumber
     \\ \label{System5} 
\end{eqnarray}
A possible solution of the above system admits the following generator 
\begin{equation}
\mathcal{X}=\left(\xi_{0} t+\xi_{1}\right) \partial_{t}+\frac{\xi_{0}}{3}(2 n-1) \partial_{a}-2 \xi_{0} R \partial_{R}+\frac{2 \xi_{0}(1-\ell)}{n} \partial_{\phi}+\left(2 \xi_{0}(1-n) \epsilon+\delta_{1}\right) \partial_{\epsilon},
\end{equation}
and the two functions 
\begin{equation}
\begin{aligned} F_{I}(R, \phi) &=\frac{\delta_{1}}{2 \xi_{0}(n-1)} R+\left[2 \xi_{0} R\right]^{n} \mathcal{F}\left(\phi+\frac{(1-n)}{\ell} \log \left[2 \xi_{0} R\right]\right), \\ F_{I I}(R, \phi) &=\frac{\delta_{1}}{2 \xi_{0}(n-1)} R+G(R) e^{k \phi}, \label{secondfunc49}\end{aligned}
\end{equation}
where $\displaystyle \mathcal{F} \big(\phi+ \frac{(1-n)}{\ell} \log[2 \xi_0 R] \big)$ is an arbitrary integration function of  $\displaystyle{\biggl(\phi+ \frac{(1-n)}{\ell} \log[2 \xi_0 R]\biggr)}$, $G(R)$ a function of the scalar curvature and $\xi_0,\xi_1, \ell, \delta_1 n, k$ constants of integration.  However, the latter function can be related to the former one  by an appropriate choice of $G(R)$. Specifically, by choosing $G(R) = G_0 R^m$, the second function becomes
\begin{equation}
F_{I I}(R, \phi) =\frac{\delta_{1}}{2 \xi_{0}(n-1)} R+G_0 R^m e^{k \phi} .
\label{FII}
\end{equation}
Also notice that Eq. \eqref{FII} is a generalization of the third function of the below system \eqref{sol non-local tegr}. Let us focus on the function $F_I$, in order to find out exact cosmological solutions.  For the sake of  simplicity, we assume the function to be linearly dependent on its argument, namely
\begin{equation}
\mathcal{F}_{1}\left(\phi+\frac{(1-n)}{\ell} \log \left[2 \xi_{0} R\right]\right) \equiv \phi+\frac{(1-n)}{\ell} \log \left[2 \xi_{0} R\right]+q,
\end{equation}
so that
\begin{equation}
F_{1}(R, \phi)=\frac{\delta_{1}}{2 \xi_{0}(n-1)} R+\left(2 \xi_{0} R\right)^{n}(q+\phi)+\left(2 \xi_{0} R\right)^{n} \frac{(1-n)}{\ell} \log \left[2 \xi_{0} R\right],
\label{genfunc}
\end{equation}
with $q$ constant. Let us note that by choosing $n=2$,  the Starobinsky gravity, non-minimally coupled to a scalar field $\phi$, is recovered. The solution of the equations of motion, for arbitrary $n$, yields three different scale factors. The first one describes a de Sitter-like expansion, which only holds for $n=3$:
\begin{equation}
\begin{array}{l} \displaystyle a(t)=a_{0} e^{\Lambda t}, \quad R(t)=-12 \Lambda^{2}, \quad \phi(t)=-\frac{1}{3}(40+3 q)-4 \Lambda t, \\ \displaystyle \epsilon(t)=576\left(2 \xi_{0}\right)^{3} \Lambda^{5} t-\frac{C_{3} e^{-3 \Lambda t}}{3 \Lambda}+\frac{\delta_{1}}{4 \xi_{0}}, \end{array}
\end{equation}
with the constraint
\begin{equation}
\Lambda=\sqrt{-\frac{1}{24 \xi_{0} e}}\qquad\left(\xi_{0}<0\right).
\end{equation}
In this case the function is further constrained by Euler-Lagrange equations, so that it reduces to
\begin{equation}
F^{(1)}_{1}(R, \phi)=\frac{\delta_{1}}{4 \xi_{0}} R+(\phi+q)\left(2 \xi_{0} R\right)^{3}-\frac{16 \xi_{0}^{3}}{\ell} R^{3} \log \left[2 \xi_{0} R\right].
\end{equation}
The second solution occurs for vanishing scalar curvature and leads to a power-law scale factor of the form:
\begin{equation}
a(t)=a_{0} t^{\frac{1}{2}}, \quad R(t)=0, \quad \phi(t)=C_{2}, \quad \epsilon(t)=\frac{\delta_{1}}{2 \xi_{0}(n-1)}-\frac{2 C_{3}}{\sqrt{t}}.
\end{equation}
Because of the vanishing Ricci scalar, in this case, the theory turns out to be equivalent to GR minimally coupled to a scalar field, described by the function
\begin{equation}
F^{(2)}_{1}(R, \phi)=\frac{\delta_{1}}{2 \xi_{0}(n-1)} R+\phi.
\end{equation}
Finally, in the last case, we obtain:
\begin{equation}
\begin{array}{l} \displaystyle a(t)=a_{0} t^{-10}, \quad R(t) \sim t^{-2}, \quad \phi(t) \sim C_{2}+ \log (t), \\ \displaystyle \epsilon(t)=\frac{\delta_{1}}{2 \xi_{0}(n-1)}+C_{3} t^{31}+c_4 \left(2 \xi_{0}\right)^{3} t^{-4},\end{array}
\end{equation}
where the corresponding function is constrained with respect to the general one in Eq. \eqref{genfunc} to
\begin{equation}
F^{(3)}_{1}(R, \phi)=\frac{\delta_{1}}{2 \xi_{0}(n-1)} R+(\phi+q)\left(2 \xi_{0} R\right)^{3}-\frac{16 \xi_{0}^{3}}{\ell} R^{3} \log \left[2 \xi_{0} R\right].
\end{equation}
Notice that, regardless of the restriction induced by the Euler-Lagrange equation solutions, the above function is still a generalization of $F^{(1)}_{1}$.

The models described in this section contain both higher order curvature invariants and local scalar fields capable of triggering both an  inflationary phase and a late-time cosmic acceleration. Respectively, it depends on the energy regime we are considering.  In order to investigate in details whether some of these solutions fit cosmological and astrophysical observations, a complete analysis is needed. Here we only provided a starting point aimed at selecting reliable non-local gravity models to solve both UV and IR issues suffered by GR and, consequently, to trace back a self-consistent cosmic history.

\section{Gauss-Bonnet Non-Local Gravity: The Case $f(\G, \Box^{-1} \G)$} \label{BoxG}
Considering the same approach as the above section,  let us now take into account a non-local IKG with action
\begin{equation}
S= \frac{1}{2 \kappa} \int \sqrt{-g} f(\G, \Box^{-1} h(\G)) d^4x \;.
\label{initial action}
\end{equation}
Eq. \eqref{initial action} depends on the Gauss-Bonnet term only and it is a generalization of the action considered in Ref. \cite{Capozziello:2008gu}. Other non-local theories of gravity, containing the Gauss--Bonnet invariant, are present in the literature, but most of them start from a well defined function which satisfies some small-scale requirements. For instance, in Ref. \cite{Elizalde:2018qbm} the authors consider an action depending on the term $\G^{n_1} \Box^{-n_{2}} \G^{n_3}$ and find cosmological solutions for some selected $n_i$. Also here, the advantage of the Noether approach is to find the functional forms of those models containing symmetries, starting from the most general action. As we will show, the Noether point symmetry condition selects five different models, and each of them leads to exact dynamics. The model in Ref. \cite{Capozziello:2008gu}, for example, represents a subcase of Eq. \eqref{initial action}, naturally provided by the Noether Symmetry Approach without any further requirement. Starting from this general approach,  it is possible to show that  the Ricci curvature scalar, and then the Hilbert-Einstein action,  can be recovered, in some particular cases,  as a limit of $f(\G)$ gravity. As mentioned in the introduction, in a FLRW background, the function $f(\G) = \sqrt{\G}$ is capable of  providing the same dynamics  as  Einstein's GR. This is directly related to the cosmological expression of the Gauss-Bonnet scalar, namely
\begin{equation}
\G = 24 \frac{\dot{a}^2 \ddot{a}}{a^3} .
\label{Gform}
\end{equation}
For this reason, the propagator $\Box^{-1} h(\G)$ can be compared to $\Box^{-1} R$ when $h(\G) = \sqrt{\G}$. The localization procedure can be pursued by introducing a local scalar field $\phi(t)$, defined as:
\begin{equation}
\Box^{-1} h(\G) := \phi(t), \;\;\;\;\;\;\; \text{so that} \;\;\;\;\;\;\; h(\G) = \Box\phi(t) \;.
\label{definition h}
\end{equation}
A new scalar field $\epsilon(t)$ can be introduced after recasting the action in terms of Lagrange multipliers, as we did in the previous case. In this way, the action \eqref{initial action} can be written as
\begin{equation}
S= \frac{1}{2 \kappa}  \int \sqrt{-g} \left\{f(\G, \phi) + \epsilon(t)(\Box \phi - h(\G)) \right\} d^4x \;.
\label{action in progress}
\end{equation}
Notice that, though $\phi$ and $\G$ are linked through the relation \eqref{definition h}, in order to write the cosmological Lagrangian and perform the Noether Symmetry Approach, they must be treated as separated fields. The minisuperspace of variables therefore turns out to be $\mathcal{Q} \equiv \{a(t), \G(t), \phi(t), \epsilon(t)\}$. Replacing the form of $\G$ in \eqref{Gform} into the action \eqref{action in progress}, we get 
\begin{equation}
S= \frac{\pi^2}{\kappa}  \int \left\{a^3 f(\G, \phi) + a^3 \epsilon(t)\left[\ddot{\phi} + 3 \frac{\dot{a}}{a} \dot{\phi} - h(\G)\right] - \lambda \left(\G - 24 \frac{\dot{a}^2 \ddot{a}}{a^3} \right) \right\}dt \;.
\end{equation}
The variation of the above action with respect to the scalar fields $\phi$ and $\epsilon$ provide the following Klein-Gordon equations
\begin{equation}
\Box \epsilon(t) = f_\phi(\G,\phi),  \; \quad \Box \phi - h(\G) = 0,
\label{KGG}
\end{equation}
respectively. The Lagrange multiplier $\lambda$ can be found in turn from the variation with respect to $\G$. After integrating out the second-derivatives carried by the d'Alembert operator, the cosmological point-like Lagrangian turns out to be
\begin{eqnarray}
\Lagr &=& a^3 \left[ f(\G,\phi) - \G f_\G(\G,\phi) - \epsilon h(\G) + \epsilon \G h_\G(\G) \right] - a^3 \dot{\phi} \dot{\epsilon} - 8 \dot{a}^3 \dot{\G} f_{\G\G}(\G,\phi)  \nonumber
\\
 &+& 8 \dot{a}^3 \dot{\epsilon} h_\G(\G) + 8 \epsilon \dot{a}^3 \dot{\G} h_{\G\G}(\G) - 8 \dot{a}^3 \dot{\phi} f_{\G\phi}(\G,\phi) \;,
\label{GB lagr}
\end{eqnarray}
where $f_\G$ and $f_\phi$ denote the derivatives of $f(\G,\phi)$ with respect to $\G$ and $\phi$, respectively. Also here, the only equations of motion providing a new contribution to the dynamics are the energy condition and the Euler-Lagrange equation with respect to the scale factor, \emph{i.e.}
\begin{eqnarray}
&& \text{E.C.} \to \dot{a} \frac{\partial \Lagr}{\partial \dot{a}} + \dot{\G} \frac{\partial \Lagr}{\partial \dot{\G}}  + \dot{\phi} \frac{\partial \Lagr}{\partial \dot{\phi}} + \dot{\epsilon} \frac{\partial \Lagr}{\partial \dot{\epsilon}}  - \Lagr = 0,
\\
&& \text{E-L} \to \frac{d}{dt} \frac{\partial \Lagr}{\partial \dot{a}} - \frac{\partial \Lagr}{\partial a} = 0.
\end{eqnarray}
They correspond to the $0,0$ and $1,1$ components of the field equations. The other three Euler-Lagrange equations yield the Klein-Gordon equations in Eq. \eqref{KGG}  and the cosmological expression of $\G$. Specifically, the entire set of equations of motion reads as:
\begin{eqnarray}
 &&  8 \dot{a} \left[2 \ddot{a} \left(-\dot{\G}f_{\G\G}(\G,\phi)-\dot{\phi}
  f_{\G \phi}(\G,\phi)+\epsilon \dot{\G} h_{\G\G}(\G)\right)+\dot{a} \left(-\ddot{\G}
  f_{\G\G}(\G,\phi) -2 \dot{\G} \dot{\phi} f_{\G\G \phi}(\G,\phi)  \right. \right. \nonumber
  \\
  && \left. -\dot{\G}^2
  f_{\G\G\G}(\G,\phi)-\ddot{\phi}f_{\G \phi}(\G,\phi)-\dot{\phi}^2
  f_{\G \phi \phi}(\G,\phi)+\ddot{\epsilon} h_\G(\G)+\epsilon \ddot{\G}
   h_{\G\G}(\G)+\epsilon \dot{\G}^2 h_{\G\G\G}(\G)\right)  \nonumber
   \\
   &&\left. +2 \dot{\epsilon} \left(\ddot{a}
   h_\G(\G)+\dot{a} \dot{\G} h_{\G\G}(\G)\right)\right]+a^2 \left[\G
  f_{\G}(\G,\phi)-f(\G,\phi)+\dot{\epsilon} \dot{\phi}+\epsilon
   \left(h(\G)-\G h_\G(\G)\right)\right]= 0, \nonumber
\end{eqnarray}
\begin{eqnarray}
&& \Box \epsilon = f_\phi(\G, \phi), \hspace{11cm} \nonumber
\end{eqnarray}
\begin{eqnarray}
\G = 24 \frac{\dot{a}^2 \ddot{a}}{a^3}, \hspace{10.5cm} \nonumber
\end{eqnarray}
\begin{eqnarray}
\Box \phi = h(\G), \hspace{10.5cm} \nonumber
\end{eqnarray}
\begin{eqnarray}
&& -a^3 \left( f(\G,\phi) - \epsilon h(\G) + \epsilon \G h_\G(\G) + \dot{\phi} \dot{\epsilon} - \G f_\G(\G,\phi) \right)  \hspace{11cm}\nonumber 
\\
&& +24 \dot{a}^3 \left( \dot{\epsilon} h_\G(\G) + \epsilon \dot{\G} h_{\G\G}(\G) - \dot{\phi} f_{\G \phi}(\G,\phi) - \dot{\G} f_{\G\G}(\G , \phi)\right) = 0 \;.
\label{EL GB}
\end{eqnarray}
Clearly, some of them are constraints. Specifically, the first equation is the Euler-Lagrange equation with respect to $a(t)$ and correspond to the $(1,1)$ component of the field equations; the second, third and fourth equations provide the Klein-Gordon equations and the cosmological expression of the Gauss-Bonnet invariant; the last equation is the energy condition and accounts for the (0,0) component of the field equations.
 Let us now apply the Noether Symmetry Approach to select the form of the two functions $f(\G, \phi)$ and $h(\G)$. In the minisuperspace considered here, the generator of the symmetry is
\begin{eqnarray}
\mathcal{X} &=& \xi(t,a,\phi,\G,\epsilon) \frac{\partial}{\partial t} + \alpha(t,a,\G,\phi,\epsilon) \frac{\partial}{\partial a}+\beta(t,a,\G,\phi,\epsilon) \frac{\partial}{\partial \G}  \nonumber
\\
&+&\gamma(t,a,\G,\phi,\epsilon)\frac{\partial}{\partial\phi}+\delta(t,a,\G,\phi,\epsilon) \frac{\partial}{\partial \epsilon},
\end{eqnarray}
where $\alpha$, $\beta$, $\gamma$ and $\delta$ are the components of the vector $\eta^i$: $\eta^i = \{\alpha, \beta, \gamma, \delta\}$. From the identity \eqref{noethid}, a system of 37 differential equations arises, though it can be reduced to five equations after neglecting linear combinations. It is \cite{Bajardi:2020mdp}:
\begin{equation}
\begin{cases}
&\displaystyle 3 \alpha a^2 f(\G,\phi) - \delta a^3 - 3 \alpha a^2 \epsilon h(\G) + \delta a^3 \G h_\G(\G)  + 3 \alpha a^2 \epsilon \G h_\G(\G) + \gamma a^3 \G \epsilon h_{\G\G}(\G) 
\\
&+ \beta a^3 f_\phi(\G,\phi) - 3 \alpha a^2 \G f_\G(\G,\phi) - \beta a^3 \G f_{\G \phi}(\G,\phi) - \gamma a^3 \G f_{\G\G}(\G,\phi) - \partial_t g 
\\
&+ a^3 \partial_t \xi \left( f(\G,\phi) - \epsilon h(\G) + \epsilon \G h_\G(\G) - \G f_\G(\G,\phi) \right) = 0
\\
\\
&\displaystyle \gamma h_{\G\G}(\G) - 3 \partial_t \xi h_\G(\G) + \partial_\epsilon \delta \; h_\G(\G) 
\\
&+ \partial_\epsilon \gamma \left(\epsilon h_{\G\G}(\G) - f_{\G\G}(\G,\phi) \right) + 3 \partial_a \alpha h_\G(\G) = 0
\\
\\
&\displaystyle f_{\G \phi}(\G,\phi) \left(3 \partial_t \xi - \partial_\phi \beta \right) + \partial_\phi \gamma \left(\epsilon h_{\G\G}(\G) - f_{\G\G}(\G,\phi) \right) \\
&- 3 \partial_a \alpha f_{\G\phi}(\G,\phi) - \beta f_{\G \phi \phi} - \gamma f_{\G \G \phi}(\G, \phi) = 0
\\
\\
&\displaystyle \beta f_{\G \G \phi}(\G,\phi) - \delta h_{\G \G}(\G) + \gamma f_{\G\G\G}(\G,\phi) - 3 \partial_t \xi f_{\G \G}(\G,\phi) + \partial_\G \gamma f_{\G\G}(\G,\phi) 
\\
&+ 3 \partial_a \alpha f_{\G \G}(\G,\phi) - \gamma \epsilon h_{\G\G\G}(\G) + 3 \partial_t \xi \epsilon h_{\G \G} - \partial_\G \gamma \epsilon h_{\G\G}(\G) - 3 \partial_a \alpha \epsilon h_{\G\G}(\G) = 0
\\
\\
&\displaystyle 3 \alpha - a \left(\partial_t \xi - \partial_\epsilon \delta - \partial_\phi \beta \right) = 0
\\
\\
&\displaystyle \alpha \equiv \alpha(a), \;\;\;\; \beta \equiv \beta(\phi), \;\;\;\; \gamma \equiv \gamma(\phi,\G,\epsilon), \;\;\;\; \delta \equiv \delta(\epsilon) ,\;\;\;\; \xi \equiv \xi(t), \;\;\;\; g \equiv g(t) \;.
\end{cases}
\label{sysGB}
\end{equation}
By solving the above system, we get five solutions:
\begin{equation}
\begin{cases}
I: \, &\mathcal{X} = (\xi_0 t + \xi_1) \partial_t + \alpha_0 a \partial_a + (\beta_0 \phi + \beta_1) \partial_\phi -4\xi_0 \G \partial_\G + \delta_0 \epsilon \partial_\epsilon,
\\
& h(\G) = h_0 \G^{\frac{1}{2} + \frac{n}{k}}, \;\;\;\;\; f(\G,\phi) =  f_0 \G^n + f_1 \G + f_2 \left(\beta_0 + \beta_1 \Box^{-1} \G^{\frac{1}{2} + \frac{n}{k}} \right)^k
\\
\\
II: \, &\mathcal{X} = (\xi_0 t + \xi_1) \partial_t + \alpha_0 a \partial_a + (\beta_0 \phi + \beta_1) \partial_\phi -4\xi_0 \G \partial_\G + (\delta_0 \epsilon + \delta_1) \partial_\epsilon,
\\
& h(\G) = h_0 \G,  \;\;\;\;\; f(\G,\phi)  = f_0 \G^n + f_1 \G + f_2 (\beta_0 \Box^{-1} \G + \beta_1)^{2n} 
\\
\\
III: \, &\mathcal{X} = (\xi_0 t + \xi_1) \partial_t + \alpha_0 a \partial_a + (\beta_0 \phi + \beta_1) \partial_\phi - 4\xi_0 \G \partial_\G + \delta_0 \epsilon \partial_\epsilon ,
\\
&h(\G) = h_0 \G^z, \;\;\;\;\; f(\G,\phi)  =  f_0 \G^n  (\beta_0 \Box^{-1} \G^z + \beta_1)^k 
\\
\\
IV: \, &\mathcal{X} = (\xi_0 t + \xi_1) \partial_t + \alpha_0 a \partial_a + (\beta_0 \phi + \beta_1) \partial_\phi -4\xi_0 \G \partial_\G + (\delta_0 \epsilon + \delta_1) \partial_\epsilon,
\\
& h(\G) = h_0 \G, \;\;\;\;\; f(\G,\phi) = f_0 \G^n (\beta_0 \Box^{-1} \G + \beta_1)^k  
\\
\\
V: \, &\mathcal{X} = (\xi_0 t + \xi_1) \partial_t + \alpha_0 a \partial_a  +\beta_1 \partial_\phi - 4 \xi_0 \G \partial_\G + \delta_0 \epsilon \partial_\epsilon,
\\
&h(\G) = h_0 \sqrt{\G}, \;\;\;\;\; f(\G,\phi) = f_0 \G^n e^{k \Box^{-1} \sqrt{\G}}    \; \,\,\,\,\, \displaystyle k \equiv \frac{\delta_0 + 4 n \xi_0}{\beta_1}
\end{cases}
\label{Solution Noether1}
\end{equation}
where $\xi_0, \, \xi_1, \, \alpha_0, \, \beta_0, \, \beta_1, \, \delta_0, \, h_0, \, f_0, \, f_1, \, f_2, \, n, \, k$ are integration constants. Some of these free parameters will be constrained by the solution of field equations. In particular, solutions II and IV do not admit any exact cosmological expression of $a(t), \phi(t), \epsilon(t)$ for any value of the free parameters; solutions I and III can be solved by setting $\beta_1 = 0$. After this imposition, the corresponding Lagrangians read 
\begin{eqnarray}
\Lagr_I &=& a^3 \left[f_0 (1-n) \G^n - h_0\left( \frac{n}{k}-\frac{1}{2}\right) \epsilon \G^{\frac{n}{k} + \frac{1}{2}} +\dot{\epsilon} \dot{\phi} \right] - 8 h_0 \left(\frac{n}{k} + \frac{1}{2} \right) \dot{a}^3 \dot{\epsilon} \G^{\frac{n}{k} - \frac{1}{2}} \nonumber
\\
&&+ 8 f_0 n(n-1) \dot{a}^3 \dot{\G} \dot{\G}^{n-2} - 8 h_0 \left(\frac{n^2}{k^2} - \frac{1}{4} \right) \epsilon \dot{a}^3 \dot{\G} \G^{\frac{n}{k} - \frac{3}{2}} +8f_1 k \dot{a}^3 \dot{\phi} \phi^{k-1}  ,
\label{sum Lagr}
\end{eqnarray}
\begin{eqnarray}
\Lagr_{III} &=& a^3 \left[f_0 (1-n) \G^n \phi^k  + h_0 (z-1) \epsilon \G^z \right] +  8 h_0 z (z-1) \epsilon \dot{a}^3 \dot{\G}  \G^{z-2}  \nonumber
\\
&& + 8 h_0 z \dot{a}^3 \dot{\epsilon}  \G^{z-1} - 8 f_0 n(n-1) \dot{a}^3 \dot{\G} \G^{n-2} \phi^k - 8 f_0 k n \dot{a}^3 \dot{\phi} \G^{n-1} \phi^{k-1} - a^3 \dot{\epsilon} \dot{\phi}. \nonumber
\\
\label{prod lagr}
\end{eqnarray}
In the former case, a power-law solution of the form
 \begin{eqnarray}
  && a(t) \sim t^{\frac{2}{3}(2n+2kz -k)}, \;\;\; \G(t) \sim t^{-4}, \;\;\; \phi(t) \sim  t^{2-4z}, \;\;\; \epsilon(t) \sim t^{2k(1-2z)} , \nonumber
  \\
  && \;\;\;\;\;\;\;\;\;\;\;\;\; f(\G, \Box^{-1} h(\G)) = f_2  \G^n (\Box^{-1} \G^z)^k, \;
 \label{EL1}
 \end{eqnarray}
occurs, which corresponds to a function $h(\G) = \G^z$ where $z$ is a real number. By setting $z = 1/2$, with the purpose to recover the GR limit, Eq. \eqref{EL1} takes the form
\begin{eqnarray}
  && a(t) \sim t^{\frac{4n}{3}}, \;\;\; \G(t) \sim t^{-4}, \;\;\; \phi(t) \sim  const., \;\;\; \epsilon(t) \sim const., \nonumber
  \\
  && \;\;\;\;\;\;\;\;\;\;\;\;\; f(\G, \Box^{-1} h(\G)) = f_2  \G^n (\Box^{-1} \sqrt{\G})^k \;.
 \end{eqnarray}
From the imposition $z = 1/2$, the scalar fields $\phi$ and $\epsilon$ turn out to be constants and the model reduces to that discussed in Ref. \cite{Bajardi:2020osh}. On the other hand, Lagrangian \eqref{prod lagr} admits a de Sitter-like vacuum solution of the form
\begin{eqnarray}
&& a(t) = a_0 e^{q t}, \;\;\; \G(t) \sim \text{const.}, \;\;\; \phi(t) \sim t, \;\;\; \epsilon(t) \sim t, \;\;\; k=1, \; n= \frac{1}{2} ,
\\
&& \;\;\;\;\;\;\;\;\;\;\;\;\; f(\G, \Box^{-1} h(\G)) = f_0 \sqrt{\G} + f_1 \G + f_2  \Box^{-1} \G + f_3 \;,
\label{EL2}
\end{eqnarray}
with $q$ a constant parameter. It is worth noticing that the above solution is the only exponential solution coming from the equations of motion. It can be obtained by setting $n=1/2$. This implies that the corresponding function $f(\G, \Box^{-1} h(\G))$ becomes
\begin{equation}
f(\G, \Box^{-1} h(\G)) = f_0 \sqrt{\G} + f_1 \G + f_2  \Box^{-1} \G + f_3.
\end{equation}
Therefore, in the case III, the only function $h(\G)$ which admits extra Noether symmetry and leads to analytic solutions of Euler-Lagrange equations is $h(\G) = \sqrt{\G}$. Notice that this is not an  imposition but it naturally arises from the solution of the equations of motion. The action, in turn, ends up to be  a sum of the terms $\sqrt{\G}$, $\Box^{-1} \G$ and $\G$. While $\sqrt{\G}$ can mimic  the scalar curvature in cosmology, $\Box^{-1} \G$ plays the role of cosmological constant. Being a topological surface term, the additive contribution of $\G$ does not provide any further contribution to the dynamics. As a consequence, the action selected by the Noether symmetry is dynamically equivalent to
\begin{equation}
S = \int \sqrt{-g} \left(f_0 R +  f_1 \Box^{-1} \G \right) d^4x \;,
\end{equation}
that is GR plus a non local term.

To conclude this section, the Lagrangian coming from the fifth function of \eqref{Solution Noether1} can be written as
\begin{eqnarray}
\Lagr_{V} &=& 2 \dot{a}^3 \left[2 \G^{-\frac{1}{2}} \dot{\epsilon} -\epsilon \G^{-\frac{3}{2}} \dot{\G} -4 f_0 n (n-1)  \G^{n-2} \dot{\G} e^{k \phi}-4 f_0 k n \G^{n-1} e^{k \phi} \dot{\phi} \right]  \nonumber 
 \\
 && -\frac{1}{2} \G^2 a^3 \left[2 \dot{\epsilon} \dot{\phi} + \epsilon \sqrt{\G}+2 f_0 (n-1) \G^n e^{k \phi}\right]
\end{eqnarray}
and, after solving the Euler-Lagrange equations, it provides
\begin{eqnarray}
&&\displaystyle a(t) = a_0 e^{q \, t}, \,\,\,\,\ \phi(t) = \sqrt{\frac{8}{3}} q\, t, \,\,\,\,\,\, \epsilon(t) \sim  e^{\sqrt{\frac{8}{3}}k q \, t},  \nonumber
\\
&& \displaystyle f(\G,\Box^{-1} \sqrt{\G}) = f_0 \G^{\frac{12 \sqrt{6}}{4k - \sqrt{6}}} e^{k \phi}, \,\,\,\,\,\,\,\,\,\, h(\G) =\sqrt{\G}.
\end{eqnarray}
In addition, the equations of motion constrain the value of the free parameter $n$, imposing a further relation between $n$ and $k$, that is: 
\begin{equation}
n = \frac{12 \sqrt{6}}{4k - \sqrt{6}}.
\end{equation}
The function is formally equivalent to that in Eq. \eqref{FII}. Also here, the de Sitter vacuum solutions are provided by the non-local terms. In other words, accelerating behaviors can be directly related to non-locality. 

\section{Teleparallel Non-Local Gravity: The Case $f(T,B. \Box^{-1} T, \Box^{-1} B)$} \label{BoxT}
Final considerations can be applied to non-locality in teleparallel gravity.
A  modification of  TEGR action, containing non-local terms of  the torsion scalar $T$ and  the boundary term $B$ can be aken into account. The above curvature-based non-local gravity is thus recovered by the application of the non-local operator $\Box^{-1}$ to the relation $R = -T+B$, namely $\Box^{-1} R =  - \Box^{-1} T + \Box^{-1} B$. The Noether symmetry condition \eqref{noethid} can be applied after introducing the Lagrange multipliers like in  Secs. \ref{BoxR} and \ref{BoxG}. Let us start from the action considered in Ref.  \cite{Bahamonde:2017sdo}, that is
\begin{equation}
S = \frac{1}{2 \kappa} \int h \left[ - T + (\tau T + \chi B) f(\Box^{-1} T, \Box^{-1} B) \right] d^4 x,
\label{actnonlocalTB}
\end{equation}
where $h$ is the determinant of tetrad fields and $\tau$, $\chi$ real constants. The Ricci scalar can be recovered by the choice $\tau = 1$, $\chi = -1$. On the other hand, when the non-local term vanishes, standard TEGR is restored. So we are considering here non-local corrections to TEGR. Notice that the action is not written in the most general form, since the contributions of the torsion and  the boundary term are not included into the function $f$. This \emph{ansatz} allows to reduce the dynamics and to restrict the minisuperspace to five dimensions. From the definitions
\begin{equation}
\Box^{-1} T \equiv \phi(t) \,\, \to \,\, T = \Box \phi(t), \qquad \Box^{-1} B \equiv \varphi(t) \,\, \to \,\, B = \Box \varphi(t),
\end{equation}
the action can be localized and a suitable Lagrangian can be obtained. It is important to notice that $T$ and $B$ are considered as two independent fields. Moreover, in analogy to the two previous sections, the two additional scalar fields $\epsilon(t)$ and $\zeta(t)$ must be considered as Lagrange multipliers. The action therefore reads:
\begin{equation}
S = \frac{1}{2 \kappa} \int h \left[- T + (\tau T+\chi B) f(\phi, \varphi) +\epsilon(\square \phi-T)+\zeta(\square \varphi-B) \right] d^4 x.
\label{dwtelep}
\end{equation} 
By varying  with respect to $\epsilon$ and $\zeta$, we get the Klein-Gordon equations
\begin{eqnarray}
&& \Box \epsilon=(\tau T+\chi B) f_\phi(\phi, \varphi),
\\
&& \Box \zeta=(\tau T+\chi B) f_\varphi(\phi, \varphi).
\end{eqnarray}
In a FLRW universe with a diagonal set of tetrad fields of the form $h = diag(1,-a(t),-a(t),-a(t))$, second derivatives of the scalar fields and of the scale factor arise into the action. Integrating out the higher-order terms, the point-like Lagrangian turns out to be:
\begin{eqnarray}
\Lagr &=& 6 a^{2} \dot{a}\left[\chi f_\phi(\phi, \varphi) \dot{\phi} + \chi f_\varphi(\phi,\varphi) \dot{\varphi} -\dot{\zeta}\right] \nonumber
\\
&+& 6 a \dot{a}^{2}\left[\epsilon + 1 - \tau f(\phi, \varphi)\right] - a^{3} \dot{\zeta} \dot{\varphi} - a^{3} \dot{\epsilon} \dot{\phi}.
\label{non-local T lagr}
\end{eqnarray}
The generator of the symmetry in the five-dimensional minisuperspace ${\cal{Q}} \equiv \{a, \phi, \varphi, \epsilon, \zeta \}$ reads:
\begin{eqnarray}
{\cal{X}} &=& \xi(a, \phi, \varphi, \epsilon, \zeta, t) \partial_t + \alpha(a, \phi, \varphi, \epsilon, \zeta, t) \partial_a + \beta(a, \phi, \varphi, \epsilon, \zeta ,t) \partial_\phi  \nonumber
\\
&&+  \gamma(a, \phi, \varphi, \epsilon, \zeta, t) \partial_\varphi + \delta(a, \phi, \varphi, \epsilon, \zeta, t) \partial_\epsilon + \theta(a, \phi, \varphi, \epsilon, \zeta, t) \partial_\zeta. \nonumber
\\
\end{eqnarray}
The application of Noether's identity \eqref{noethid} to the above Lagrangian yields a system of 43 partial differential equations, but only 18 of them are linearly independent \cite{Bahamonde:2017sdo}:
\begin{equation}
\begin{cases} 
\displaystyle a \partial_t \theta -6 \chi f_{\varphi} \partial_t \alpha =0
\\
\displaystyle a\partial_t \delta -6 \chi f_{\phi} \partial_t \alpha =0
\\
\displaystyle 6 \partial_t \alpha +a \partial_t \gamma =0
\\
\displaystyle a \left(\partial_\varphi \beta + \partial_\epsilon \theta \right)- 6 \chi f_\varphi \partial_\epsilon \alpha =0
\\
\displaystyle a\left(\partial_\zeta \beta + \partial_\epsilon \gamma \right)+6 \partial_\epsilon \alpha =0
\\
\displaystyle 6 \chi f_{\phi}\partial_\phi \alpha -a \partial_\phi \delta =0
\\
\displaystyle 6 \chi f_\varphi \partial_\varphi \alpha -a \partial_\varphi \theta =0
\\
\displaystyle 6 \partial_\zeta \alpha +a \partial_\zeta \varphi =0
\\
\displaystyle -6 a^{2}\left(\chi f_{\phi} \partial_t \beta +\chi f_{\varphi} \partial_t \gamma - \partial_t \theta\right) +12 a (\tau f - \epsilon - 1) \partial_t \alpha = 0 
\\
\displaystyle 6 a\left(\chi f_\varphi \partial_\epsilon \gamma- \partial_\epsilon \theta \right)+12(\epsilon+1-\tau f) \partial_\epsilon \alpha -a^{2} \partial_a \beta=0
\\
\displaystyle 6 \chi\left(f_{\varphi} \partial_\varphi \alpha +f_{\varphi} \partial_\phi \alpha \right) - a\left( \partial_\varphi \delta + \partial_\phi \theta \right)=0
\\
\displaystyle 6\left(\chi f_{\varphi} \partial_\zeta \alpha -\partial_\phi \alpha \right)-a\left(\partial_\phi \gamma +\partial_\zeta \delta \right)=0
\\
\displaystyle 6 \chi f_{\varphi} \partial_\epsilon \alpha-a\left(\partial_\phi \beta + \partial_\epsilon \delta -\partial_t \xi \right)-3 \alpha =0
\\
\displaystyle 6 \chi f_{\varphi} \partial_\zeta \alpha - 6 \partial_\varphi \alpha - 3 \alpha -a\left(\partial_\varphi \gamma +\partial_\zeta \theta - \partial_t \xi \right)=0
\\
\displaystyle 6 a\left(\chi f_{\varphi} \partial_\phi \beta +\chi f_{\phi \phi} \beta +\chi f_{\phi} \partial_a \alpha +\chi f_{\varphi} \partial_\phi \gamma  +\chi f_{\phi \varphi} \gamma -\chi\partial_t \xi f_{\phi}-\partial_\phi \theta \right)
\\
\displaystyle+ 12 \chi f_{\phi}\alpha +12(\epsilon + 1-\tau f) \partial_\phi \alpha -a^{2} \partial_a \delta = 0
\\
\displaystyle 6 a\left(\chi f_{\phi} \partial_\zeta \beta + \chi f_{\varphi} \partial_\zeta \gamma -\partial_a \alpha -\partial_\zeta \theta + \partial_t \xi \right)
\\
\displaystyle+12(\epsilon+1-\tau f) \partial_\zeta \alpha -12 \alpha -a^{2} \partial_a \gamma =0
\\
\displaystyle 6 a\left[\chi f_{\phi} \partial_\varphi \beta +\chi f_{\phi \varphi} \beta +\chi f_\varphi \left(\partial_a \alpha +\partial_\varphi \gamma - \partial_t \xi \right) +\chi f_{\varphi \varphi} \gamma - \partial_\varphi \theta \right] 
\\
\displaystyle+12 \chi f_{\varphi} \alpha + 12(\epsilon+1-\tau f) \partial_\varphi \alpha -a^{2} \partial_a \theta =0
\\
\displaystyle a\left[\chi a f_{\phi}\partial_a \beta- \tau f_{\phi} \beta +\chi a f_{\varphi} \partial_a \gamma -\tau f_{\varphi} \gamma -2 \tau f \partial_a \alpha +\tau \partial_t \xi f + 2\epsilon \partial_a \alpha  \right.
\\
\left. +2 \partial_a \alpha +\delta -a \partial_a \theta -(\epsilon+1) \partial_t \xi \right) +(\epsilon+1-\tau f) \alpha =0
\\
\displaystyle \xi = \xi(t), \,\,\,\, \beta = \beta(a, \phi, \varphi, \zeta, t).
\end{cases}
\label{non-local tegr system}
\end{equation}
Due to the dimension of  minisuperspace, the system turns out to be over determined. Therefore, here we outline some solutions of cosmological interest, writing the infinitesimal generators and the corresponding functions containing symmetries. They are:
\begin{equation}
\begin{cases}
\displaystyle  \begin{aligned} &{\cal{X}}_1=  \left(\xi_0 t + \xi_1 \right) \partial_{t}+\frac{ \alpha_0}{3} a \partial_{a} + \left[\beta_0+ \beta_1(6 \ln a+\phi)\right] \partial_{\phi} \\& +\left[\gamma_0 + \gamma_1(6 \ln a+\varphi) + \gamma_2 \right] \partial_{\varphi}+(\xi_0 - \alpha_0) \epsilon \partial_{\epsilon}  +\left[\left(\xi_0 - \alpha_0 - \gamma_1 \right) \zeta - \beta_1 \epsilon + \theta_0 \right] \partial_{\zeta},  \end{aligned} 
\\
\displaystyle f(\phi, \varphi)=\frac{1}{\tau}+ f_0 \exp \left\{n \left(\beta_1 \varphi-\gamma_1 \phi\right)\right\}, \quad n = \frac{\xi_0 - \alpha_0}{\beta_1 \gamma_0 - \beta_0 \gamma_1 + \beta_1 \gamma_2}
\\
\\
\displaystyle  \begin{aligned} {\cal{X}}_2 =&\left(\xi_0 t + \xi_1 \right) \partial_{t}+\frac{\alpha_0}{3} a \partial_{a} +\left[\beta_0+\beta_1(6 \ln a+\varphi)\right] \partial_{\phi} \\ & +\gamma_0 \partial_{\varphi}+\left(\delta_0 +\delta_1 \epsilon \right) \partial_{\epsilon} +\left(\delta_1 \zeta- \beta_1 \epsilon + \theta_0 \right) \partial_{\zeta},  \end{aligned}
\\
\displaystyle f(\phi, \varphi)=\frac{1}{\tau} \left(1 - \frac{\delta_0}{\delta_1} \right) + f_0  e^{\frac{\delta_1}{\gamma_0} \varphi}
\\
\\
\displaystyle  {\cal{X}}_3 =\left(\xi_0 t + \xi_1 \right) \partial_{t}-\frac{\alpha_0}{3} a \partial_{a} + \beta_0 \partial_{\phi}+\left(\delta_0 + \delta_1 \epsilon \right) \partial_{\epsilon} ,
\\
\displaystyle f(\phi) = f_0 e^{\frac{\beta_0}{\delta_1} \phi} - \frac{\beta_0}{\delta_0}+1, \quad \text{with} \quad \chi = 0, \, \zeta = 0,\, \tau = 1
\\
\\
\displaystyle {\cal{X}}_4=\left(\xi_0 t + \xi_1 \right) \partial_{t}-\frac{\alpha_0}{3} a \partial_{a} + \beta_0 \partial_{\phi}+\delta_0 \partial_{\epsilon} ,
\\
\displaystyle f(\phi) = f_1 + \frac{\beta_0}{\delta_0} \phi, \quad \text{with} \quad \chi = 0, \, \zeta = 0, \, \tau = 1,
\end{cases}
\label{sol non-local tegr}
\end{equation}
where $\xi_0, \alpha_0, \alpha_1, \beta_0, \beta_1, \gamma_0, \gamma_1, \gamma_2, \delta_0, \delta_1, \theta_0, f_0, f_1, n$ are  integration constants. It is worth remarking that we neglected those solutions which do not uniquely determine the form of $f(\phi, \varphi)$ (see \cite{Bahamonde:2017sdo} for details). The last Noether system solution is the only one not providing  exact cosmological solutions, thus, in what follows, we investigate the cosmological implications of the first three cases.

The first solution yields the Lagrangian
\begin{eqnarray}
\Lagr = &a& \left\{ 6 (\epsilon -f_0 \tau e^{n( \beta_1 \varphi - \gamma_1 \phi)}) \dot{a}^2 - a^2 \dot{\epsilon} \dot{\phi} \right. \nonumber
\\
&+& \left. 6 a \dot{a}  \left[\chi f_0 n e^{n( \beta_1 \varphi - \gamma_1 \phi)}  (\beta_1 \dot{\varphi}-\gamma_1 \dot{\phi} ) - \dot{\zeta}  \right]-  a^2 \dot{\varphi} \dot{\zeta} \right\}, \nonumber
\\ \label{lagrtegr}
\end{eqnarray}
whose cosmological solutions are
\begin{eqnarray}
&& a(t)=a_0 e^{H_{0} t}, \qquad  \phi(t)=-2 H_{0} t, \qquad  \nonumber \varphi(t)=-6 H_{0} t,
\\
&&\epsilon(t) = \epsilon_0 e^{-3H_0 t (1 + c_1)} - \epsilon_1 e^{-3 H_{0} t} , \qquad \zeta(t)= \zeta_0 e^{-3H_0 t (1 + c_1)} - \zeta_1 e^{-3 H_{0} t}, \nonumber
\\
\end{eqnarray}
and 
\begin{eqnarray}
&& a(t)=a_0 t^{p}, \qquad \phi(t)=\frac{6 p^{2} \ln (t-3 p t)}{1-3 p}, \qquad \varphi(t)=-6 p \ln t, \nonumber
\\
&& \epsilon(t) = \epsilon_0 t^{2-3p} + \epsilon_1 t^{1-3p}, \qquad \zeta(t) = \zeta_0 t^{2-3p} + \zeta_1 t^{1-3p}, \nonumber
\\
\end{eqnarray}
with $H_0$, $p$ constants. In the former case, the following relation occurs:
\begin{equation}
\xi_0 = \alpha_0 - \frac{(3 \beta_1 - 2 \gamma_1) \tau - 3 \gamma_1 \chi}{2 \chi(2 \beta_1 - \gamma_1)^2},
\end{equation}
so that $n$ is constrained to
\begin{equation}
n = - \frac{(3 \beta_1 - 2 \gamma_1) \tau - 3 \gamma_1 \chi}{2 \chi(2 \beta_1 - \gamma_1)^2}.
\end{equation}
However, it is worth noticing that both exponential and power-law solutions are admitted by the equations of motion in non-local TEGR. As a consequence, the latter can potentially describe all the cosmic epochs, from the early to the late times. Moreover, standard local TEGR is recovered for $\tau = \chi = 0$, namely when $ n = 0$, while modified $f(R)$ gravity is restored as soon as $T$ and $B$ are combined through the relation in Eq. \eqref{R = -T+B}.

The second solution of the Noether system \eqref{sol non-local tegr} can be related to the first one for appropriate combinations of the free parameters. For this reason, it will not be discussed in this section (see \cite{Bahamonde:2017sdo} for details). A Lagrangian  corresponding to the third function of \eqref{sol non-local tegr}  can be also obtained as a limit of Lagrangian \eqref{non-local T lagr} and reads as:
\begin{equation}
\mathcal{L}=-6 a \dot{a}^{2}\left(f_0 e^{\frac{\beta_0}{\delta_1} \phi} -\epsilon -1\right)-a^{3} \dot{\epsilon} \dot{\phi}.
\end{equation}
According to the constraints provided by the Noether Approach, we set $\chi = 0, \zeta = 0, \tau = 1$. In this way, the equations of motion can be solved analytically providing
\begin{eqnarray}
&& a(t)=e^{H_{0} t}, \quad \phi(t)=-2 H_{0} t, \quad \epsilon(t)=e^{-3 H_{0} t}\left[f_0\left(3 H_{0} t+1\right)-\frac{\epsilon_1}{3 H_{0}}\right] -1, \nonumber
\\
&& f(\phi, \varphi) = f_0 e^{\frac{\beta_0}{\delta_1} \phi},
\label{sol111}
\end{eqnarray} 
and 
\begin{eqnarray}
&& a(t) = t^p, \quad \phi(t)=\frac{6 p^{2} \log (t-3 p t)}{1-3 p}, \quad \epsilon(t)= f_0 (1-3p)^{3(1-p)} t^{2-3p} \frac{\epsilon_0 t^{1-3 p}}{1-3 p}-1, \nonumber
\\
&& f(\phi)=f_0 e^{\frac{\left(9 p^{2}-9 p+2\right) \phi}{6 p^{2}}}.
\label{sol222}
\end{eqnarray}
A further constraint must be imposed in the search of exact solutions, namely $\delta_0 = \beta_0$. By means of this imposition, the Euler-Lagrange equations and the energy condition can be solved analytically, yielding the exponential and  power-law scale factors in Eqs. \eqref{sol111} and \eqref{sol222}. 

Finally, it is worth analyzing a further solution, coming from the introduction of a new scalar field $\psi$, defined as:
\begin{equation}
\psi \equiv - \phi + \varphi = - \Box^{-1} T + \Box^{-1} B = \Box^{-1}(-T+B) = \Box^{-1} R.
\label{redefpsi}
\end{equation}
In this way, $f(R)$ gravity can be straightforwardly recovered and the minisuperspace can be reduced to 
$\mathcal{Q} \equiv \{a, \psi, \epsilon, \zeta \}$.  As a consequence of this assumption, the Noether system admits two solutions, namely
\begin{equation}
\begin{cases}
\displaystyle \begin{aligned} {\cal{X}}_5 =\left(\xi_0 t + \xi_1 \right) \partial_{t}-\frac{\alpha_0}{3} a \partial_{a} - 2 c_0 \partial_{\psi}+\left(\delta_0 + \delta_1 \epsilon \right) \partial_{\epsilon}, \end{aligned}
\\
\displaystyle f(\psi) = -1+\frac{\delta_0}{\delta_{1}}+f_0 e^{-\frac{\delta_{1}}{\beta_{0}} \psi}
\\
\\
\displaystyle  \begin{aligned} {\cal{X}}_6 =\left(\xi_0 t + \xi_1 \right) \partial_{t}-\frac{\alpha_0}{3} a \partial_{a} - 2 c_0 \partial_{\psi}+\delta_0 \partial_{\epsilon},\end{aligned}
\\
\displaystyle f(\psi)  = f_1+\frac{\delta_0}{2 \beta_0} \psi.
\end{cases}
\end{equation}
Both functions arise after a redefinition of the scalar fields $\phi$ and $\varphi$ in  Eq. \eqref{redefpsi} and thus can be obtained from solutions \eqref{sol non-local tegr} with an appropriate change of variables.The function therefore turns out to be equivalent to $f(\Box^{-1} R)$, so that the action \eqref{NL:22} can be recovered. As a consequence, also the cosmological solutions are the same as in the case of curvature based non-local gravity  discussed in Sec. \ref{BoxR}.

Notice that in Sec. \ref{BoxR},  three different solutions occurred after the application of the Noether identity. This difference is due to the  form of the initial actions, where the torsion scalar and the boundary term appear linearly. 

\section{ Spherical Symmetry in Non-local gravity. The Case of S2 Star Orbit}
\label{sectS2}
In the previous sections, we dealt with non-local gravity in a cosmological background. Here we apply the same method but in a  spherically symmetric spacetime given by the interval
\begin{equation}
ds^2 = e^{\nu(r,t)} dt^2 - e^{\lambda(r,t)} dr^2 - r^2 d\Omega^2.
\end{equation} 
In order to get analytic solutions by the Noether symmetries  and obtain a suitable set of equations of motion, we take into account  the spherically symmetric  action 
\begin{equation}
S = \frac{1}{2 \kappa} \int \sqrt{-g} \, \Big\{R[1 + f(\phi)] + \varepsilon(r,t) (\Box \phi - R)\Big\} d^4x\,,
\label{dwaction}
\end{equation}
where the auxiliary field $\phi$ to localize the dynamics has been defined as above.

In this way, the configuration space (the minisuperspace) is  ($\mathcal{Q} \equiv \{\nu, \lambda, \phi, \varepsilon\}$) because the scalar curvature can be written explicitly in terms of the metric potentials $e^\nu$ and $e^\lambda$. Notice that the imposition 
\begin{equation}
\nu \equiv \nu(r) = - \lambda(r),
\end{equation}
cannot be applied \emph{a priori}, since the relation between $\lambda$ and $\nu$, related to the Schwarzschild solution, occurs after the field equations resolution. 

The purpose of this section is to show that  the free parameters of the point-like spherically symmetric Lagrangian, selected by the existence of Noether symmetries, can be  astronomically constrained taking into account the data coming from the S2 star, orbiting around SgrA$^*$. The constraint will be performed  in the weak field limit taking into account Newtonian potential correction related to non-locality.

It is worth noticing that the above action is written in terms of the scalar fields $\varepsilon(r,t)$ and $\phi(r,t)$, by means of the same localization procedure as the previous sections. The Klein-Gordon equations with respect to $\phi$ and $\varepsilon$ read, respectively:
\begin{equation}
\frac{\delta S}{\delta \phi} = 0 \, \to \, \Box \varepsilon = - R f_\phi, \qquad \frac{\delta S}{\delta \varepsilon} = 0 \, \to \, \Box \phi =  R ,
\end{equation}
while the variation with respect to the metric yields the field equations 
\begin{equation}
[1+f(\phi) - \varepsilon] G_{\mu \nu}= \left(D_{\mu} D_{\nu}-g_{\mu \nu} \square\right) \, f(\phi) - \frac{1}{2} g_{\mu \nu} D_\alpha \varepsilon D^\alpha \phi + D_\mu \varepsilon D_\nu \phi.
\end{equation}
Before deriving the point-like Lagrangian, it is worth noticing that the metric depends both on the radial coordinate $r$ and on the coordinate time $t$. Therefore, the infinitesimal generator $\xi^\mu$ turns out to be a two-vector with components $\xi^t$ and $\xi^r$. Considering the form of the d'Alembert operator in spherical symmetry and integrating out the second derivatives, the canonical Lagrangian reads as:
\begin{eqnarray}
\Lagr(r,\nu,\lambda) &=& e^{-\frac{1}{2}(\lambda+\nu)}\left[-e^{\nu} r^{2} \nu_{r} \phi_{r} f_\phi(\phi)+e^{\lambda} r^{2} \lambda_{t} \phi_{t} f_\phi(\phi) \right. \nonumber
\\
&&- 2 e^{\nu} f(\phi)\left(e^{\lambda}+r \lambda_{r}-1\right)-2 e^{\lambda+\nu}+2 e^{\nu}+e^{\nu} r^{2} \varepsilon_{r} \phi_{r} +e^{\nu} r^{2} \nu_{r} \varepsilon_{r} \nonumber
\\
&&\left.  -e^{\lambda} r^{2} \varepsilon_{t} \phi_{t}-e^{\lambda} r^{2} \lambda_{t} \varepsilon_{t} +2 e^{\nu} \varepsilon \left(e^{\lambda}+r \lambda_{r}-1\right)-2 e^{\nu} r \lambda_{r}\right],
\label{lagrsph}
\end{eqnarray}
whose symmetry generator is
\begin{equation}
\mathcal{X} = 	\xi^t \partial_t + \xi^r \partial_r + \alpha \partial_\nu + \beta \partial_\lambda + \gamma \partial_\phi + \delta \partial_\varepsilon.
\end{equation}
The application of Noether's Theorem to the point-like Lagrangian \eqref{lagrsph} selects the two models \cite{Dialektopoulos:2018iph}
\begin{equation}
\begin{cases}
\displaystyle \mathcal{X} = (\xi_0 t + \xi^t(r)) \partial_t - 2\xi_0 \partial_\nu + (\gamma_0 + 2 \xi_0) \partial_\phi + \delta_0  (\gamma_0 + 2 \xi_0)  \partial_\varepsilon,
\\
\displaystyle f(\phi) = \delta_0 \phi + f_1
\\
\\
\displaystyle  \mathcal{X} = (\xi_0 t + \xi^r(r)) \partial_t - \frac{\xi_1}{2} r \partial_r - (2\xi_0 + \xi_1) \partial_\nu + \gamma_0 \partial_\phi + \xi_1  (\varepsilon - \delta_0 -1)  \partial_\varepsilon,
\\
\displaystyle f(\phi) = \delta_0 + f_1 e^{\frac{\gamma_0}{\xi_1} \phi}.
\end{cases}
\label{noeth spherical non-local}
\end{equation}

In order to investigate the weak-field limit, we restrict the space of solutions to a subclass where the Birkhoff theorem holds, namely where both $\nu$ and $\lambda$ are independent of time. This assumption is reasonable, as a first approximation, since in the weak-field limit a static and spherically symmetric spacetime is a solution of the field equations and reasonably represents the dynamics around SgrA$^*$. In this perspective, let us therefore consider the line element
\begin{equation}
ds^2 = A(r) dt^2 - B(r) \, dr^2 - r^2 d\Omega^2,
\end{equation}
and study the Post-Newtonian limit of the theory, with the aim to constrain the non-local action by the observations coming from S2 star orbit. To this purpose, let us expand the $g_{00}$ component of the metric up to the sixth order and the  $g_{11}$ component up to the fourth:
\begin{equation}
g_{00} \sim \mathcal{O}(6),\;\; g_{0 i} \sim \mathcal{O}(5)\;\; \text { and } \;\; g_{i j} \sim \mathcal{O}(4).
\end{equation}
The  expansions provide 
\begin{equation}
\begin{cases}
\displaystyle A(r)=1+\frac{1}{c^{2}} \Phi(r)^{(2)}+\frac{1}{c^{4}} \Phi(r)^{(4)}+\frac{1}{c^{6}} \Phi(r)^{(6)}+\mathcal{O}(8)
\\
\displaystyle B(r)=1+\frac{1}{c^{2}} \Psi(r)^{(2)}+\frac{1}{c^{4}} \Psi(r)^{(4)}+\mathcal{O}(6)
\\
\displaystyle \phi(r)=\phi_{0}+\frac{1}{c^{2}} \phi(r)^{(2)}+\frac{1}{c^{4}} \phi(r)^{(4)}+\frac{1}{c^{6}} \phi(r)^{(6)}+\mathcal{O}(8)
\\
\displaystyle \varepsilon(r)=\varepsilon_0 +\frac{1}{c^{2}} \varepsilon(r)^{(2)}+\frac{1}{c^{4}} \varepsilon(r)^{(4)}+\frac{1}{c^{6}}\varepsilon(r)^{(6)}+\mathcal{O}(8),
\end{cases}
\label{perturb}
\end{equation}
with $\Phi(r)$ and $\Psi(r)$ being the Newtonian potential coming from the  expansion of $g_{00}$ and $g_{11}$, respectively, while $\phi_0$ and $\varepsilon_0$ are constants. Considering the second function of \eqref{noeth spherical non-local} and setting $\delta_0 = f_1 =1$ and $\gamma_0 = \xi_1$, the Klein-Gordon equations together with the field equations yield the system:
\begin{equation}
\begin{split}
&2 B^{2}\left(-\varepsilon+e^{\phi}+2\right)+r B^{\prime}\left(-2 \varepsilon-r \varepsilon^{\prime}+r e^{\phi} \phi^{\prime}+2 e^{\phi}+4\right)  
\\
&-B\left[-2 \varepsilon+2\left(-r^{2} \varepsilon^{\prime \prime}+r^{2} e^{\phi} \phi^{\prime \prime}+r^{2} e^{\phi}\left(\phi^{\prime}\right)^{2}+2 r e^{\phi} \phi^{\prime}+e^{\phi}+2\right)+r \varepsilon^{\prime}\left(r \phi^{\prime}-4\right)\right]=0, 
\\
\\
& r A^{\prime}\left(-2 \varepsilon-r \varepsilon^{\prime}+r e^{\phi} \phi^{\prime}+2 e^{\phi}+4\right) 
\\
&-A\left[2 B\left(-\varepsilon+e^{\phi}+2\right)+2 \varepsilon+r^{2} \varepsilon^{\prime} \phi^{\prime} +4 r \varepsilon^{\prime}-4 r e^{\phi} \phi^{\prime}-2 e^{\phi}-4\right]=0,
\\ 
\\
& A^{2}\left[-4 B^{2} e^{\phi}+r B^{\prime}\left(r \varepsilon^{\prime}-4 e^{\phi}\right)+B\left(-2 r^{2} \varepsilon^{\prime \prime}-4 r \varepsilon^{\prime}+4 e^{\phi}\right)\right] 
\\
& + B r^{2}\left(-e^{\phi}\right)\left(A^{\prime}\right)^{2}+A r\left\{B\left[2 r e^{\phi} A^{\prime \prime}+A^{\prime}\left(4 e^{\phi}-r \varepsilon^{\prime}\right)\right]-r e^{\phi} A^{\prime} B^{\prime}\right\}=0,
\\ 
\\
& A^{2}\left[-4 B^{2}-r B^{\prime}\left(r \phi^{\prime}+4\right)+2 B\left(r^{2} \phi^{\prime \prime}+2 r \phi^{\prime}+2\right)\right]+B\left(-r^{2}\right)\left(A^{\prime}\right)^{2} 
\\
&+A r\left\{B\left[2 r A^{\prime \prime}+A^{\prime}\left(r \phi^{\prime}+4\right)\right]-r A^{\prime} B^{\prime}\right\}=0,
\end{split}
\end{equation}
where the prime denotes the derivative with respect to $r$. The solution of the above equations can be written in terms of the effective gravitational constant $G_{eff} = G_N \phi_c$, with $\phi_c$ a real constant. Replacing the perturbations \eqref{perturb} into the above system, we get
\begin{eqnarray}
A(r)&=&1-\frac{2 G_{N} M \phi_{c}}{c^{2} r}+\frac{G_{N}^{2} M^{2}}{c^{4} r^{2}}\left[\frac{14}{9} \phi_{c}^{2}+\frac{18 r_{\varepsilon}-11 r_{\phi}}{6 r_{\varepsilon} r_{\phi}} r\right] \nonumber
\\
&& -\frac{G_{N}^{3} M^{3}}{c^{6} r^{3}}\left[\frac{50 r_{\varepsilon}-7 r_{\phi}}{12 r_{\varepsilon} r_{\phi}} \phi_{c} r+\frac{16 \phi_{c}^{3}}{27}-\frac{r^{2}\left(2 r_{\varepsilon}^{2}-r_{\phi}^{2}\right)}{r_{\varepsilon}^{2} r_{\phi}^{2}}\right], \nonumber
\\  \nonumber
\\
B(r)&=&1+\frac{2 G_{N} M \phi_{c}}{3 c^{2} r}+\frac{G_{N}^{2} M^{2}}{c^{4} r^{2}}\left[\frac{2 \phi_{c}^{2}}{9}+\left(\frac{3}{2 r_{\varepsilon}}-\frac{1}{r_{\phi}}\right) r\right], \nonumber
\\ \nonumber
\\
\phi(r)&=&\frac{4 G_{N} M \phi_{c}}{3 c^{2} r}-\frac{G_{N}^{2} M^{2}}{c^{4} r^{2}}\left[\left(\frac{11}{6 r_{\varepsilon}}+\frac{1}{r_{\phi}}\right) r-\frac{2 \phi_{c}^{2}}{9}\right] \nonumber
\\
&& -\frac{G_{N}^{3} M^{3}}{c^{6} r^{3}}\left[\frac{r^{2}}{r_{\phi}^{2}}-\left(\frac{25}{12 r_{\varepsilon}}-\frac{7}{6 r_{\phi}}\right) \phi_{c} r-\frac{4 \phi_{c}^{3}}{81}\right], \nonumber
\\ \nonumber
\\
\varepsilon(r)&=&1+\frac{G_{N}^{2} M^{2}}{c^{4} r^{2}}\left[\frac{2 \phi_{c}^{2}}{3}-\left(\frac{13}{6 r_{\varepsilon}}-\frac{1}{r_{\phi}}\right) r\right]  \nonumber
\\
&&+\frac{G_{N}^{3} M^{3}}{c^{6} r^{3}}\left[\frac{20 \phi_{c}^{3}}{27}-\left(\frac{1}{r_{\varepsilon}^{2}}-\frac{1}{r_{\phi}^{2}}\right) r^{2}-\left(\frac{131}{36 r_{\varepsilon}}+\frac{1}{6 r_{\phi}}\right) \phi_{c} r\right], \nonumber 
\end{eqnarray} 
where $r_\varepsilon$ and $r_\phi$ are length scales, related to the scalar fields $\varepsilon$ and $\phi$, which can be addressed to the non-locality. The experimental observations of S2 star orbit can be used to constrain the free parameters $\phi_c$, $r_\phi$ and $r_\varepsilon$. From the second-order expansion of the potential $\Phi$, that is
\begin{equation}
\Phi^{(2)}(r)=-\frac{2 G_{N} M}{r} \phi_{c},
\end{equation}
it is possible to infer that the constant $\phi_c$ must be equal to $1$. In this way, the Newtonian potential can be recovered as a particular limit. The other two free parameters must be constrained by the two-body simulation of S2, orbiting around SgrA$^*$.

Let $ \mu = M \cdot m_s/(M+m_s)$ be the reduced mass, with $M$ being the mass of SgrA$^*$ and $m_S$ the mass of S2. 

According to the numeric data provided in Ref. \cite{Gillessen:2008qv}, we vary the values of $r_\varepsilon$ and $r_\phi$, searching for those solutions which provides a lower $\chi^2$ than the Keplerian orbit ($\chi^2 \sim 1.89$). Following Refs. \cite{Dialektopoulos:2018iph, Jovanovic:2019bdy}, the reduced $\chi^2$ can be written as:
\begin{equation}
\chi^{2}=\frac{1}{2 N-\nu} \sum_{i=1}^{N}\left[\left(\frac{x_{i}^{o}-x_{i}^{c}}{\sigma_{x i}}\right)^{2}+\left(\frac{y_{i}^{o}-y_{i}^{c}}{\sigma_{y i}}\right)^{2}\right],
\end{equation}
where ($x^o_i, y^o_i$) and ($x^c_i$, $y^c_i$) are the observed and the theoretical apparent positions, respectively. $N$ is the number of observations, $\nu$ the number of initial conditions, $\sigma_{x i}$ and $\sigma_{y i}$ the uncertainties of the observed positions. The graphs in Fig. 2 shows the $\chi^2$ in different regions of the parameter space $r_\phi - r_\varepsilon$.
\begin{center}
\centering
\includegraphics[width=.40\textwidth]{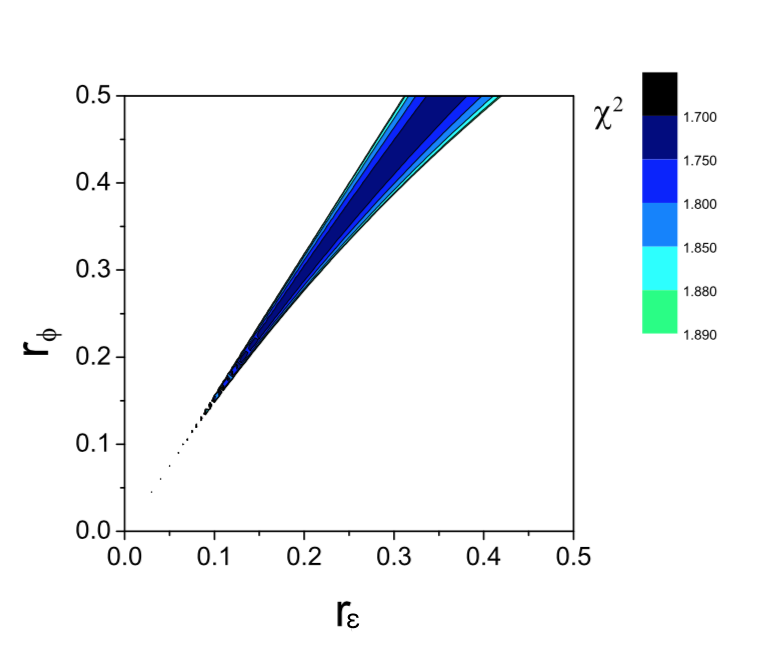}
\end{center}
\begin{center}
\centering
\includegraphics[width=.7\textwidth]{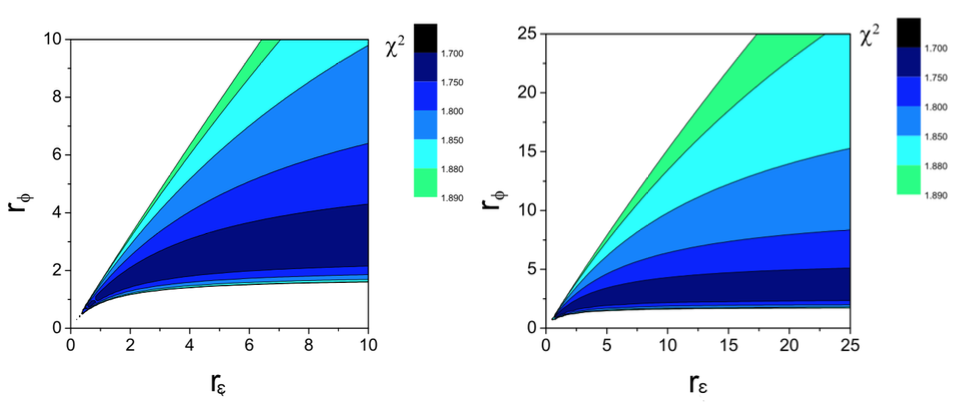}
\end{center}
\begin{center}
\centering
\includegraphics[width=.7\textwidth]{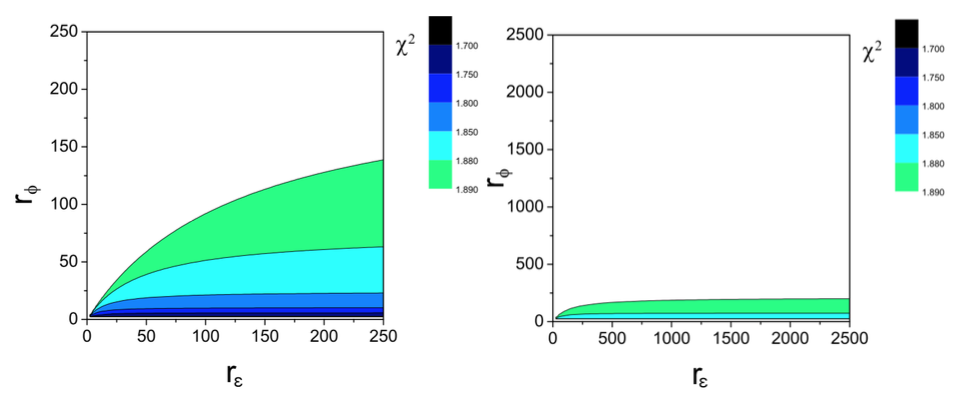}
\\ Figure 2: \emph{Maps of the reduced $\chi^2$ in different regions of the parameter space. Figures report the lengths $r_\phi$ and $r_\varepsilon$ in AU for all those simulated orbits of S2 star giving a lower $\chi^2$ than the Keplerian orbit. The colors are darker when the fit is better, namely when the $\chi^2$ is lower.}
\end{center}

From the above graphs, we notice that the best value for $r_\phi$ is comprehended in the range $r_\phi \sim 0.1 - 2.5$ AU. However, the analysis of the reduced $\chi^2$ can only constrain the length $r_\phi$, since the parameter $r_\varepsilon$ is related to the non-dynamical scalar field $\varepsilon$, which is just a mathematical tool aimed at localizing the theory.

The potential energy in the weak-field limit can be calculated by means of the expansion of the potential $\Phi$:
\begin{eqnarray}
&&\Phi^{(2)}(r)=-\frac{2 G_{N} M}{r} \phi_{c}, \nonumber
\\
&& \Phi^{(4)}(r)=\frac{G_{N}^{2} M^{2}}{r^{2}}\left[\frac{14}{9} \phi_{c}^{2}+\frac{18 r_{\varepsilon}-11 r_{\phi}}{6 r_{\varepsilon} r_{\phi}} r\right], \nonumber
\\
&& \Phi^{(6)}(r)=\frac{G_{N}^{3} M^{3}}{r^{3}}\left[\frac{7 r_{\phi}-50 r_{\varepsilon}}{12 r_{\varepsilon} r_{\phi}} \phi_{c} r-\frac{16 \phi_{c}^{3}}{27}+\frac{2 r_{\varepsilon}^{2}-r_{\phi}^{2}}{r_{\varepsilon}^{2} r_{\phi}^{2}} r^{2}\right], 
\end{eqnarray}
so that the energy reads
\begin{equation}
\begin{aligned} U_{N L}=&-\frac{G_{N} M}{r} \phi_{c}+\frac{G_{N}^{2} M^{2}}{2 c^{2} r^{2}}\left[\frac{14}{9} \phi_{c}^{2}+\frac{18 r_{\varepsilon}-11 r_{\phi}}{6 r_{\varepsilon} r_{\phi}} r\right] \\ &+\frac{G_{N}^{3} M^{3}}{2 c^{4} r^{3}}\left[\frac{7 r_{\phi}-50 r_{\varepsilon}}{12 r_{\varepsilon} r_{\phi}} \phi_{c} r-\frac{16 \phi_{c}^{3}}{27}+\frac{2 r_{\varepsilon}^{2}-r_{\phi}^{2}}{r_{\varepsilon}^{2} r_{\phi}^{2}} r^{2}\right] .\end{aligned}
\label{energy}
\end{equation}
Eq. \eqref{energy} contains non-local corrections related to the function $f(\phi)$, from which it is possible to compute the precession per orbital period (see Fig. 3).
\begin{center}
\centering
\includegraphics[width=0.9\textwidth]{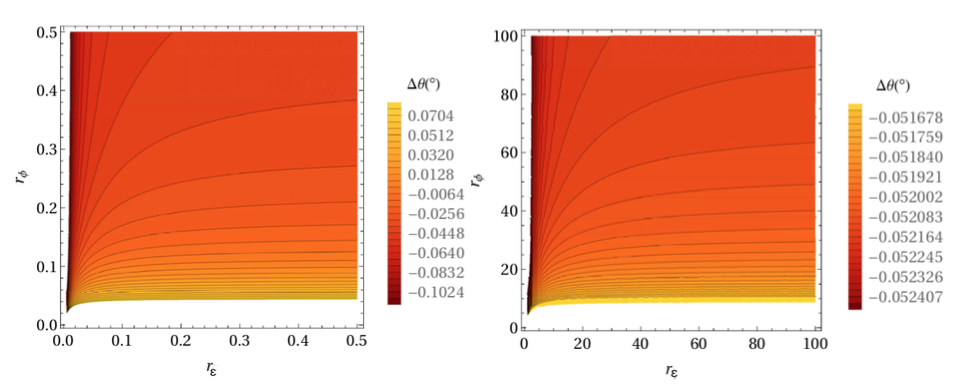}
\\ Figure 3: \emph{Precession per orbital period in two different regions of the parameter space. Darker colors refer to lower values of the precession angles.}
\end{center}

Finally, considering the values $r_\phi \sim 1.2$ AU and $r_\varepsilon \sim 1.1$ AU, which correspond to $\chi^2 \sim 1.72$, the comparison between the Kelplerian orbit of S2 star and the non-local orbit yields the plot in Figs. 4-5.

\begin{center}
\centering
\includegraphics[width=0.4\textwidth]{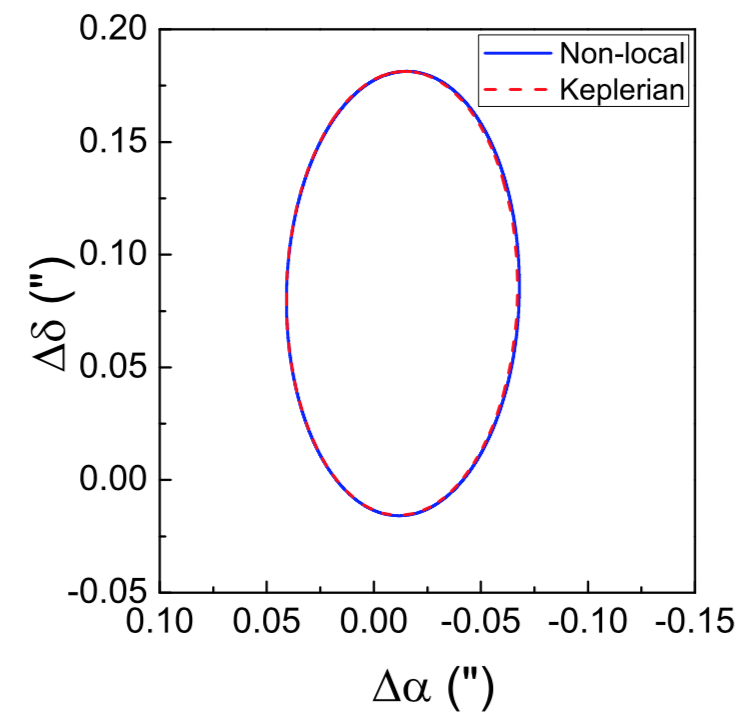}
\\ Figure 4: \emph{The Figure shows the comparison between the Keplerian and the non-local orbit. Specifically, the red dashed line represents the Keplerian orbit, while the blue solid line represents the orbit obtained in the non-local theory. The values of the parameters are those providing the lowest $\chi^2$. $\Delta \alpha$ and $\Delta \delta$ are the coordinates of S2 star.}
\end{center}
\begin{center}
\centering
\includegraphics[width=0.4\textwidth]{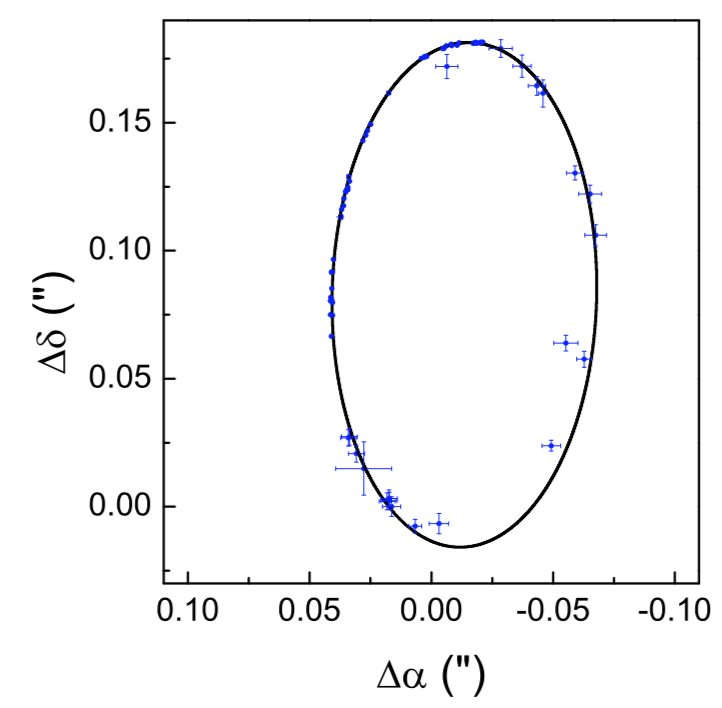}
\\ Figure 5: \emph{The Figure shows the  best fit orbit in non-local gravity, with parameters $r_\phi$ and $r_\varepsilon$ minimizing the $\chi^2$: $r_\phi \sim 1.2$ AU and $r_\varepsilon \sim 1.1$ AU. The fitted orbit is compared to the data provided by NTT/VLT observations. $\Delta \alpha$ and $\Delta \delta$ are the coordinates of S2 star.}
\end{center}

Figs. 4 and 5 show that some regions are in agreement with observations even more than the Keplerian case. The selected non-local action is then reduced such that the scalar degrees of freedom $r_\phi$ and $r_\varepsilon$ perfectly fit astrometric data. The results point out that corrections coming from non-local gravity effects can be compared with data. This is just an example but, in principle, other astrophysical scales can be fitted against data to highlight possible non-local effects. See also \cite{Salzano}.

\section{Conclusions and Perspectives} \label{Concl}

We examined some features of non-locality  in Physics, mainly focusing on non-local theories of gravity. Specifically, we considered three different classes of IKGs and selected the corresponding actions by the Noether Symmetry Approach.  In  Sec. \ref{BoxR} a curvature based non-local action  $F(R, \Box^{-1} R)$ is considered. The existence of symmetries allows to find out exact cosmological solutions that naturally address accelerated behavior without any fine tuning of parameters.  It is worth noticing that the models selected by symmetries agree with prescriptions of unitarity and super-renormalizability of some effective theory of quantum gravity (see  
\cite{Modesto:2017hzl, Tomboulis:2015esa, Modesto:2013ioa, Modesto:2011kw,Briscese:2012ys} for details). This fact indicates that Noether Symmetry Approach could be a criterion to select physically motivated theories (see also \cite{SergeyMF}).

From this point of view, a future perspective for cosmology coming from  non-local theories of gravity is to  constrain  free parameters by observation. An example in this research line is reported in \cite{Bahamonde:2017bps}.

In Secs. \ref{BoxG} and \ref{BoxT} non-local theories of gravity containing the Gauss-Bonnet term and the torsion scalar are considered, respectively. Also in these cases, after localizing the  theory by means of an additional scalar field, we adopted Noether Symmetry Approach to select viable models.   The resulting scale factors turn out to present exponential and  power-law behavior. This means that non-local terms can not only reproduce all the main cosmic epochs but also  play the role of cosmological constant even without introducing any dark energy. It is worth noticing that curvature, Gauss-Bonnet and torsion representations can be always related each other giving a comprehensive approach to the possible extensions of GR. In all particular representations of exteded gravity, non-locality plays an important role with respect to the issues of Dark Energy, Dark Matter and Large Scale Structure. 

As an example of this statement, 
 in Sec. \ref{sectS2}, we considered a minimal non-local extension of GR in spherical symmetry. The related model can be constrained by astrometric data of S2 star orbit. This analysis can be pursued by varying the length scales which minimize the reduced $\chi^2$, until the latter fits the experimental observations better than the standard Keplerian orbits. This fact, in our opinion, is particularly relevant because points out that non-local effects can be directly investigated by observations at galactic scales. The spherically symmetric analysis, however, is pursued only in $f(R, \Box^{-1} R)$ gravity. Nevertheless, under given limits, both $f(T, B, \Box^{-1} T, \Box^{-1} B)$ and $f(\G, \Box^{-1} \G)$ can reproduce similar results, in the sense that the free parameters coming from the Noether Symmetry Approach can be constrained by the S2 star orbit. In the former case, we expect to recover exactly the same results as in $f(R, \Box^{-1} R)$ gravity, if the torsion and the boundary term are combined according to Eq. \eqref{R = -T+B}. This is due to the fact that the action in Eq. \eqref{dwtelep} is formally equivalent to the Deser-Woodard action \eqref{dwaction}. However, if the boundary term is not considered, symmetries provided by $f(T, \Box^{-1} T)$ are generally different and the spherically symmetric analysis may not provide the same solutions. Nonetheless, GR is still recovered in the absence of the non-local term and the free parameters arising from the Noether approach can be constrained to fit the astrometric data. Similar considerations also hold in the case of $f(\G, \Box^{-1} \G)$ gravity, since, in some backgrounds, the Gauss--Bonnet invariant can reproduce the dynamical behavior of the scalar curvature even without imposing the GR limit  \cite{Bajardi:2020osh, Bajardi:2019zzs}. In future works, we aim to study these non-local theories in a spherically symmetric space-time, comparing the results with those provided in Sec. \ref{sectS2} and constraining the free parameters by  observations.

\section*{Acknowledgments}

The Authors acknowledge the support of {\it Istituto Nazionale di Fisica Nucleare} (INFN) ({\it iniziative specifiche} GINGER, MOONLIGHT2, and QGSKY). 

\appendix

\section{The Noether Symmetry Approach} \label{noeth}
Let us  briefly summarize the main aspects of Noether Symmetry Approach, adopted in Secs. \ref{BoxR}, \ref{BoxG}, \ref{BoxT} and \ref{sectS2}. Basic foundations and applications of the general prescription can be found \emph{e.g.} in \cite{Rugg,Urban:2020lfk, Bajardi:2021tul, Capozziello:2007wc, Capozziello:2008ch, Hussain:2011wa}. Besides the ordinary application of the well known Noether theorem, we start by assuming that the Lagrangian is invariant under some transformation involving coordinates $x^\mu$ and fields $\phi^i$, namely:
\begin{equation}
\left\{\begin{array}{l}\tilde{x}^{\mu}=x^{\mu}+\epsilon \xi^{\mu}\left(x^{\mu}, \phi^{i}\right)+O\left(\epsilon^{2}\right) \\ \tilde{\phi}^{i}=\phi^{i}+\epsilon \eta^{i}\left(x^{\mu}, \phi^{i}\right)+O\left(\epsilon^{2}\right),\end{array}\right. 
\label{trans}
\end{equation}
with $\xi^\mu$ and $\eta^i$ being the infinitesimal generators of the symmetry transformation. The total generator of the transformation is therefore
\begin{equation}
\mathcal{X} = \xi^{\mu} \partial_{\mu}+\eta^{i} \frac{\partial}{\partial \phi^{i}},
\label{gengen}
\end{equation}
and it is possible to show that if the condition
\begin{equation}
\left[\xi^{\mu} \partial_{\mu}+\eta^{i} \frac{\partial}{\partial \phi^{i}}+\left(\partial_{\mu} \eta^{i}-\partial_{\mu} \phi^{i} \partial_{\nu} \xi^{\nu}\right) \frac{\partial}{\partial\left(\partial_{\mu} \phi^{i}\right)}+\partial_{\mu} \xi^{\mu}\right] \mathscr{L}=\partial_{\mu} g^{\mu},
\label{noethid}
\end{equation}
holds, then the quantity
\begin{equation}
j^{\mu}=-\frac{\partial \mathscr{L}}{\partial\left(\partial_{\mu} \phi^{i}\right)} \eta^{i}+\frac{\partial \mathscr{L}}{\partial\left(\partial_{\mu} \phi^{i}\right)} \partial_{\nu} \phi^{i} \xi^{\nu}-\mathscr{L} \xi^{\mu}+g^{\mu},
\end{equation}
is a first integral of motion. Here $g^\mu$ stands for a generic function of coordinates and fields, called \emph{Gauge Function}. By the definition 
\begin{equation}
X^{[1]}=\xi^{\mu} \partial_{\mu}+\eta^{i} \frac{\partial}{\partial \phi^{i}}+\left(\partial_{\mu} \eta^{i}-\partial_{\mu} \phi^{i} \partial_{\nu} \xi^{\nu}\right) \frac{\partial}{\partial\left(\partial_{\mu} \phi^{i}\right)}
\label{x1}\,,
\end{equation}
 the identity \eqref{noethid} can be written as
\begin{equation}
X^{[1]} \mathscr{L}+\partial_{\mu} \xi^{\mu} \mathscr{L} =\partial_{\mu} g^{\mu}.
\label{idx1}
\end{equation}
The vector $X^{[1]}$ is called \emph{first prolongation of Noether's vector}. In cosmology, where the variables are only time-dependent, the definition \eqref{x1} and the identity \eqref{idx1} take the form
\begin{equation}
X^{[1]}=\xi \frac{\partial}{\partial t}+\eta^{i} \frac{\partial}{\partial \phi^{i}}+\eta^{i[1]} \frac{\partial}{\partial \dot{\phi}^{i}}=\xi \frac{\partial}{\partial t}+\eta^{i} \frac{\partial}{\partial \phi^{i}}+\left(\dot{\eta}^{i}-\dot{\phi}^{i} \dot{\xi}\right) \frac{\partial}{\partial \dot{\phi}^{i}},
\end{equation}
and
\begin{equation}
X^{[1]} \mathcal{L}+\dot{\xi} \mathcal{L}=\dot{g}\left(t, \phi^{i}\right),
\end{equation}
respectively. If the variables in the Lagrangian are only a functions of the radius, the above equations become
\begin{equation}
X^{[1]}=\xi\left(r, \phi^{i}\right) \partial_{r}+\eta^{i}\left(r, \phi^{i}\right) \frac{\partial}{\partial \phi^{i}}+\left[\partial_{r} \eta^{i}\left(r, \phi^{i}\right)-\partial_{r} \phi^{i} \partial_{r} \xi\left(r, \phi^{i}\right)\right] \frac{\partial}{\partial\left(\partial_{r} \phi^{i}\right)},
\end{equation}
and
\begin{equation}
X^{[1]} \mathcal{L}+\partial_{r} \xi\left(r, \phi^{i}\right) \mathcal{L}=\partial_{r} g\left(r, \phi^{i}\right).
\end{equation}
For internal symmetries, where the infinitesimal generator $\xi^\mu$ vanishes, the condition \eqref{noethid} takes the form
\begin{equation}
\label{symm}
\left[\eta^{i} \frac{\partial}{\partial \phi^{i}}+  \partial_{\mu} \eta^{i} \frac{\partial}{\partial\left(\partial_{\mu} \phi^{i}\right)} \right] \mathscr{L}= \partial_{\mu} g^{\mu}.
\end{equation}
By arbitrarily setting $g^\mu = 0$, the above equation can be recast in terms of the vanishing Lie derivative of the Lagrangian along the flux of the generator $\mathcal{X}$.  Clearly, Eq.\eqref{symm} gives rise to a first integral into the Euler-Lagrange equations. This fact allows to reduce the dynamics and, eventually, solve it. In all cases presented above, the procedure allows also to find out physically motivated models.

\end{document}